\documentclass[aps,prx,reprint,superscriptaddress,footinbib,floatfix,longbibliography]{revtex4-2}

\usepackage{graphicx}
\usepackage{bm}
\usepackage{amssymb}
\usepackage{amsbsy}
\usepackage{amsmath}
\usepackage{mathtools}
\usepackage{xcolor}
\usepackage{stmaryrd}
\usepackage{mathrsfs}
\usepackage[mathscr]{euscript}
\usepackage{tikz}
\usetikzlibrary{shapes}
\usepackage{cancel}
\usepackage[normalem]{ulem}
\usepackage{hyperref}

\begin{document}

\title{How important are fluctuations in the treatment of internal friction in polymers?}

\author{R. Kailasham}
\affiliation{IITB-Monash Research Academy, Indian Institute of Technology Bombay, Mumbai, Maharashtra -  400076, India}
\affiliation{Department of Chemistry, Indian Institute of Technology Bombay, Mumbai, Maharashtra -  400076, India}
\affiliation{Department of Chemical Engineering, Monash University,
Melbourne, VIC 3800, Australia}
\author{Rajarshi Chakrabarti}
\email{rajarshi@chem.iitb.ac.in}
\affiliation{Department of Chemistry, Indian Institute of Technology Bombay, Mumbai, Maharashtra -  400076, India}
\author{J. Ravi Prakash}
\email{ravi.jagadeeshan@monash.edu}
\affiliation{Department of Chemical Engineering, Monash University,
Melbourne, VIC 3800, Australia}

\begin{abstract}
The Rouse model with internal friction (RIF), a widely used theoretical framework to interpret the effects of internal friction on conformational transitions in biomolecules, is shown to be an approximate treatment that is based on preaveraging internal friction. By comparison with Brownian dynamics simulations of an exact coarse-grained model that incorporates fluctuations in internal friction, the accuracy of the preaveraged model predictions is examined both at and away from equilibrium. While the two models predict intrachain autocorrelations that approach each other for long enough chain segments, they differ in their predictions for shorter segments. Furthermore, the two models differ qualitatively in their predictions for the chain extension and viscosity in shear flow, which is taken to represent a prototypical out-of-equilibrium condition.
\end{abstract}

\maketitle

\section{\label{sec:intro}Introduction}
Macromolecules in solution experience an additional mode of dissipation or friction due to intramolecular interactions, over and above the solvent drag, which resists their conformational reconfiguration~\cite{kuhn1945,Booij1970,degennes,Ansari1992,Kailasham2018,Kailasham2020}. This additional mode of dissipation termed as \textit{internal friction}~\cite{Bird1987b,ravibook,Cellmer2008,DeSancho2014,Samanta2014,Samanta2016165} (IV), has been known to significantly affect the conformational dynamics~\cite{Qiu2004,Wensley2010,Hagen2010385,Borgia2012,Soranno201217800,Soranno2017,Socol2019} of chains and the rheology~\cite{Manke1988,Dasbach1992,Gerhardt1994} of polymer solutions. In force-spectroscopic measurements on polysaccharides or condensed DNA~\cite{Alexander-Katz2009,Schulz20154565,Murayama2007,Khatri20071825}, the dissipation associated with stretching the molecules is much greater than that can be attributed to solvent friction alone. An important class of intramolecular chemical reactions~\cite{Guerin2012} relies on the formation of a loop between distant segments of a polymer chain, and internal friction has been shown to affect the looping and reconfiguration times in polymer molecules~\cite{Cheng2013,Samanta2014,Samanta2016165,Soranno2017}. The most widely used theoretical framework for the interpretation of internal friction effects~\cite{Soranno201217800,ja211494h,Ameseder2018,Soranno2018} is the Rouse model with internal friction~\cite{Khatri20076770} (RIF) and its variants~\cite{Cheng2013,Samanta2014,Samanta2016165}, which modify the standard continuum Rouse model to include an additional frictional force that resists changes in the curvature of the space-curve representing the polymer molecule. While these models remain preferred due to their analytical tractability, their accuracy, from a theoretical standpoint, has so far not been examined.  We show that the RIF model neglects fluctuations in the internal friction force and is essentially equivalent to a preaveraged treatment. We have recently developed an exact numerical solution to the Rouse model with fluctuating internal friction [Ref.~\citenum{kailasham2021rouse}, under review], and have used it to estimate linear-viscoelastic and steady-shear viscometric functions with the help of Brownian dynamics (BD) simulations.  In this paper, the exact model is used to test the accuracy of the RIF model, by comparing the predictions of the two models for quantities both at and away from equilibrium.

The effect of internal friction on the dynamics of protein reconfiguration is commonly quantified experimentally by tagging the molecule with fluorescent donor-acceptor pairs along their contour length, and extracting a characteristic reconfiguration time from the autocorrelation of the fluorescence signal~\cite{Soranno2017,Soranno2018}. Analytical and simulation estimates of the reconfiguration time are based on the autocorrelation of the vector that connects the tagged monomers along the chain~\cite{Cheng2013,Samanta2014,Samanta2016165}. We examine this equilibrium property with a view to quantifying the influence of fluctuations in internal friction. In order to examine the importance of fluctuations away from equilibrium, a polymer molecule subjected to simple shear flow is considered as a prototypical out-of-equilibrium process. Several biological processes, such as ciliary and flagellar oscillations in microorganisms~\cite{Poirier2002,Mondal2020,Nandagiri2020}, are driven by the hydrolysis of ATP molecules, and the contribution from internal friction in these far-from-equilibrium processes is seen to outweigh hydrodynamic drag by nearly an order of magnitude. Protein molecules such as hormones and antibodies are commonly subjected to shear flow during various stages of bioprocessing~\cite{Bekard2011}. The conformational dynamics of these molecules in flow directly affects their structure and function, which further adds relevance to the study of the dynamic response of such polymers to shear flow.

We find that the equilibrium predictions made by the preaveraged model and the one with fluctuations differ at small separations between the tagged monomers along the polymer backbone, with the difference diminishing with an increase in the inter-tag separation. However, in the presence of shear flow, the two model predictions differ starkly and qualitatively: the preaveraged model predicts values for the chain extension and viscosity that are identical to the standard Rouse model, with the internal friction parameter only affecting the transient phase that precedes the attainment of steady state. Contrarily, exact BD simulations which account for fluctuations in IV establish that both the transients and the steady-state values are modified by internal friction~\cite{kailasham2021rouse}.

The rest of this paper is organized as follows. In Section~\ref{sec:sol_disc_rif}, the discrete version of the RIF model and the pertinent expressions for observables at equilibrium and in flow are presented. In Section~\ref{sec:rouse_preav}, the formal equivalence between the discrete RIF model, and the bead-spring-dashpot chain model with preaveraged internal friction, derived using the principles of polymer kinetic theory (PKT), is established. Results on the effect of fluctuations at equilibrium [Sec.~\ref{sec:fluc_eqb}] and in flow [Sec.~\ref{sec:fluc_shear_flow}] are presented next, followed by concluding remarks in Sec.~\ref{sec:concl_fluc_imp}. Appendix~\ref{sec:app_a} outlines the main steps for the derivation of the autocorrelation of interbead connector vectors in a discrete RIF chain at equilibrium. The equivalence between the discrete and continuum versions of the RIF model, and the derivation of an analytical expression for the mean-squared end-to-end distance of the discrete RIF model in shear flow are presented in Appendices~\ref{sec:d_cont_rif_equiv} and~\ref{sec:derv_shear_flow}, respectively. The derivation of the governing stochastic differential equations for the bead-spring-dashpot chain with preaveraged internal friction is presented in Appendix~\ref{sec:derv_preav_sde}, and a semi-analytical solution to the model is derived in Appendix~\ref{sec:equivalence}. Detailed steps for the derivation of the appropriate stress tensor expression for chains with preaveraged internal friction have been presented in Appendix~\ref{sec:derv_stress_tensor}.

\section{\label{sec:sol_disc_rif} Solution of discrete RIF model}

The standard RIF model is in the continuous chain limit, but it is convenient to work with a discrete model for the sake of comparison with simulations. The discrete RIF model has been solved using normal-mode analysis in Appendix~\ref{sec:app_a}, and as demonstrated in Figs.~\ref{fig:sum_satur} and~\ref{fig:disc_cont_compare} in Appendix~\ref{sec:d_cont_rif_equiv}, the discrete model is identical to the continuous chain RIF model as the number of beads, $N_{\text{b}} \gg1$. For ease of exposition, only the key results of the discrete RIF model are presented here. In this model, the beads, each of radius $a$ and suspended in a solvent of viscosity $\eta_{\text{s}}$, are located at positions $\{\bm{r}_1,\bm{r}_2,\cdots,\bm{r}_{N_{\text{b}}}\}$, and connected by Hookean springs of stiffness $H$ in parallel with dashpots that have a damping coefficient $(K/3)$. The reason for the appearance of the factor of 3 becomes clear below. The dashpots provide a resistive force that is proportional to the relative velocity between adjacent beads, and the force due to internal friction on a bead $\mu$ not at the chain ends is given by
\begin{align}\label{eq:iv_force_rif}
\bm{F}^{\text{(IV),RIF}}_{\mu}&=\left({K}/{3}\right)\left(\dot{\bm{r}}_{\mu+1}-\dot{\bm{r}}_{\mu}\right)-\left({K}/{3}\right)\left(\dot{\bm{r}}_{\mu}-\dot{\bm{r}}_{\mu-1}\right)\nonumber\\[2pt]
&=(K/3)\left[\dot{\bm{r}}_{\mu+1}-2\dot{\bm{r}}_{\mu}-\dot{\bm{r}}_{\mu-1}\right]
\end{align}
where $\dot{\bm{r}}_{\mu}=d\bm{r}_{\mu}/dt$. The overdamped Langevin equation for the time evolution of $\bm{r}_{\mu}$, is given by
\begin{equation}\label{eq:bead_govern_with_shear}
\dfrac{d\bm{r}_{\mu}}{dt}=-\left(\dfrac{H}{\zeta}+\dfrac{\varphi}{3}\dfrac{d}{dt}\right)\sum_{\nu=1}^{N_{\text{b}}}A^{\text{(R)}}_{\mu \nu}\bm{r}_{\nu}+\boldsymbol{\kappa}\cdot\bm{r}_{\mu}+\boldsymbol{\xi}_\mu(t)
\end{equation}
where $\zeta=6\pi\eta_{\text{s}}a$ is the bead friction coefficient, $A^{\text{(R)}}_{\mu \nu}$ are elements of the Rouse matrix~\cite{Bird1987b} (defined in Eq.~(\ref{eq:rouse_mat_def})), and $\boldsymbol{\kappa}$ represents the flow-field, which is $\bm{0}$ in the absence of flow, and
\begin{equation}\label{eq:kappa_shear_def}
\boldsymbol{\kappa} = 
\begin{pmatrix}
0& \dot{\gamma} & {0}\\
0& 0& 0\\
{0} & 0 & 0 
\end{pmatrix}
\end{equation}
for simple shear flow, where $\dot{\gamma}$ denotes the shear rate. The internal friction parameter, $\varphi\equiv\,K/\zeta$, is the ratio of the damping coefficient of the dashpot to the bead friction coefficient. The characteristic length- and time-scales in the coarse-grained models discussed above are defined to be $l_{H}=\sqrt{k_BT/H}$ and $\lambda_{H}=\zeta/4H$, respectively, where $k_B$ is Boltzmann's constant and $T$ the absolute temperature. Scaled dimensionless variables are denoted with an asterisk as superscript. The moments of the noise term, $\boldsymbol{\xi}_\mu(t)$, are not specified in real space, but rather in normal-mode space. Essentially, Eq.~(\ref{eq:bead_govern_with_shear}) can be solved (as detailed in  Appendix~\ref{sec:app_a}) by first using the eigenvectors, $a_{p}=4\sin^2\left({p\pi}/{2N_{\text{b}}}\right);\,p\in[0,(N_{\text{b}}-1)]$, of the Rouse matrix for projecting the bead positions into normal-mode space, followed by the assumption that the noise term in normal mode space is white, so as to satisfy the requirements of equipartition. This treatment results in a solution which is similar to the standard Rouse model, with a renormalization of the mode relaxation times. The expression for the normalized dimensionless autocorrelation of the interbead connector vector, $\bm{R}^{*}_{\mu\nu}\equiv\bm{r}^{*}_{\nu}-\bm{r}^{*}_{\mu}$, for the discrete RIF model at equilibrium ($\boldsymbol{\kappa}=\bm{0}$) is then
\begin{widetext}
\begin{equation}\label{eq:mu_nu_correl_in_paper}
\begin{split}
\dfrac{\left<\boldsymbol{R}^{*}_{\mu\nu}(0)\cdot\boldsymbol{R}^{*}_{\mu\nu}(t^*)\right>}{\left<\boldsymbol{R}^{*2}_{\mu\nu}(0)\right>}=\left[\dfrac{2}{N_{\mathrm{b}}|\nu-\mu|}\right]\,\sum_{p=1}^{N_{\text{b}}-1}&\left\{\cos\left[\left(\nu-\dfrac{1}{2}\right)\dfrac{p\pi}{N_{\mathrm{b}}}\right]-\cos\left[\left(\mu-\dfrac{1}{2}\right)\dfrac{p\pi}{N_{\mathrm{b}}}\right]\right\}^2\left(\dfrac{1}{a_p}\right)\exp\left[-\left(\dfrac{3a_p}{3+\varphi a_p}\right)\dfrac{t^*}{4}\right]
\end{split}
\end{equation}
and the dimensionless, normalized autocorrelation of the end-to-end vector, $\bm{R}^{*}_{\text{E}}=\bm{r}^{*}_{N_{\text{b}}}-\bm{r}^{*}_{1}$, may therefore be rewritten as
\begin{equation}\label{eq:disc_exp_nm_dimless_main}
\begin{split}
\dfrac{\left<\bm{R}^{*}_{\text{E}}(0)\cdot\bm{R}^{*}_{\text{E}}(t^*)\right>}{\left<\bm{R}^{*2}_{\text{E}}(0)\right>}&=\left[\dfrac{8}{N_{\mathrm{b}}\left(N_{\text{b}}-1\right)}\right]\,\sum_{p:\text{odd}}^{N_{\text{b}}-1}\cos^2\left(\dfrac{p\pi}{2N_{\mathrm{b}}}\right)\left(\dfrac{1}{a_p}\right)\exp\left[-\left(\dfrac{3a_p}{3+\varphi a_p}\right)\dfrac{t^*}{4}\right]
\end{split}
\end{equation}
In addition to the expressions above for the autocorrelation of the interbead and end-to-end connector vectors at equilibrium, we have also derived (as described in Appendix~\ref{sec:derv_shear_flow}) an expression for the transient evolution of the dimensionless mean-squared end-to-end vector in shear flow scaled by its equilibrium value, $\left<\bm{R}^{*2}_{\text{E}}(t^{*})\right>/\left<\bm{R}^{*2}_{\text{E}}\right>_{\text{eq}}$, 
 \begin{equation}\label{eq:nmlzd_trans_re_t}
\begin{split}
&\dfrac{\left<\bm{R}_{\text{E}}^{*2}(t^*)\right>}{\left<\bm{R}^{*2}_{\text{E}}\right>_{\text{eq}}}=\left[\dfrac{8}{N_{\mathrm{b}}(N_{\mathrm{b}}-1)}\right]\,\sum_{p:\text{odd}}^{N_{\text{b}}-1}\left(\dfrac{1}{a_p}\right)\cos^2\left(\dfrac{p\pi}{2N_{\mathrm{b}}}\right)\\[5pt]
&\times\Biggl\{1+\dfrac{8\left(\lambda_{{H}}\dot{\gamma}\right)^2}{3a_p^2}\Biggl[1-\Biggl(\exp\left[{-\left(\dfrac{3a_p}{3+\varphi a_p}\right)\dfrac{t^{*}}{2}}\right]\left[1+\left(\dfrac{3a_p}{3+\varphi a_p}\right)\dfrac{t^{*}}{2}\right]\Biggr)\Biggr]\Biggr\}
\end{split}
\end{equation}
where $\left<\bm{R}^{*2}_{\text{E}}\right>_{\text{eq}}=3\left(N_{\text{b}}-1\right)$ is the mean-squared value for the end-to-end vector at equilibrium. The steady state value is obtained by taking the limit $t^{*}\to\infty$ in Eq.~(\ref{eq:nmlzd_trans_re_t}) to give
 \begin{equation}\label{eq:nmlzd_steady_re_t}
\begin{split}
&\dfrac{\left<\bm{R}_{\text{E}}^{*2}\right>}{\left<\bm{R}^{*2}_{\text{E}}\right>_{\text{eq}}}=\left[\dfrac{8}{N_{\mathrm{b}}(N_{\mathrm{b}}-1)}\right]\,\sum_{p:\text{odd}}^{N_{\text{b}}-1}\left(\dfrac{1}{a_p}\right)\cos^2\left(\dfrac{p\pi}{2N_{\mathrm{b}}}\right)\Biggl\{1+\dfrac{8\left(\lambda_{{H}}\dot{\gamma}\right)^2}{3a_p^2}\Biggr\}
\end{split}
\end{equation}
which is independent of the internal friction parameter. It is worth noting that a solution for the continuum RIF chain subjected to shear flow has so far not been derived. In the limit of large $N_{\text{b}}$, Eq.~(\ref{eq:nmlzd_steady_re_t}) is found to agree with the result for a continuum model of a Rouse chain in shear flow~\cite{Bhattacharyya2012}.
\end{widetext}

\begin{figure*}[pthb]
\begin{center}
\begin{tabular}{c c}
\includegraphics[width=2.7in,height=!]{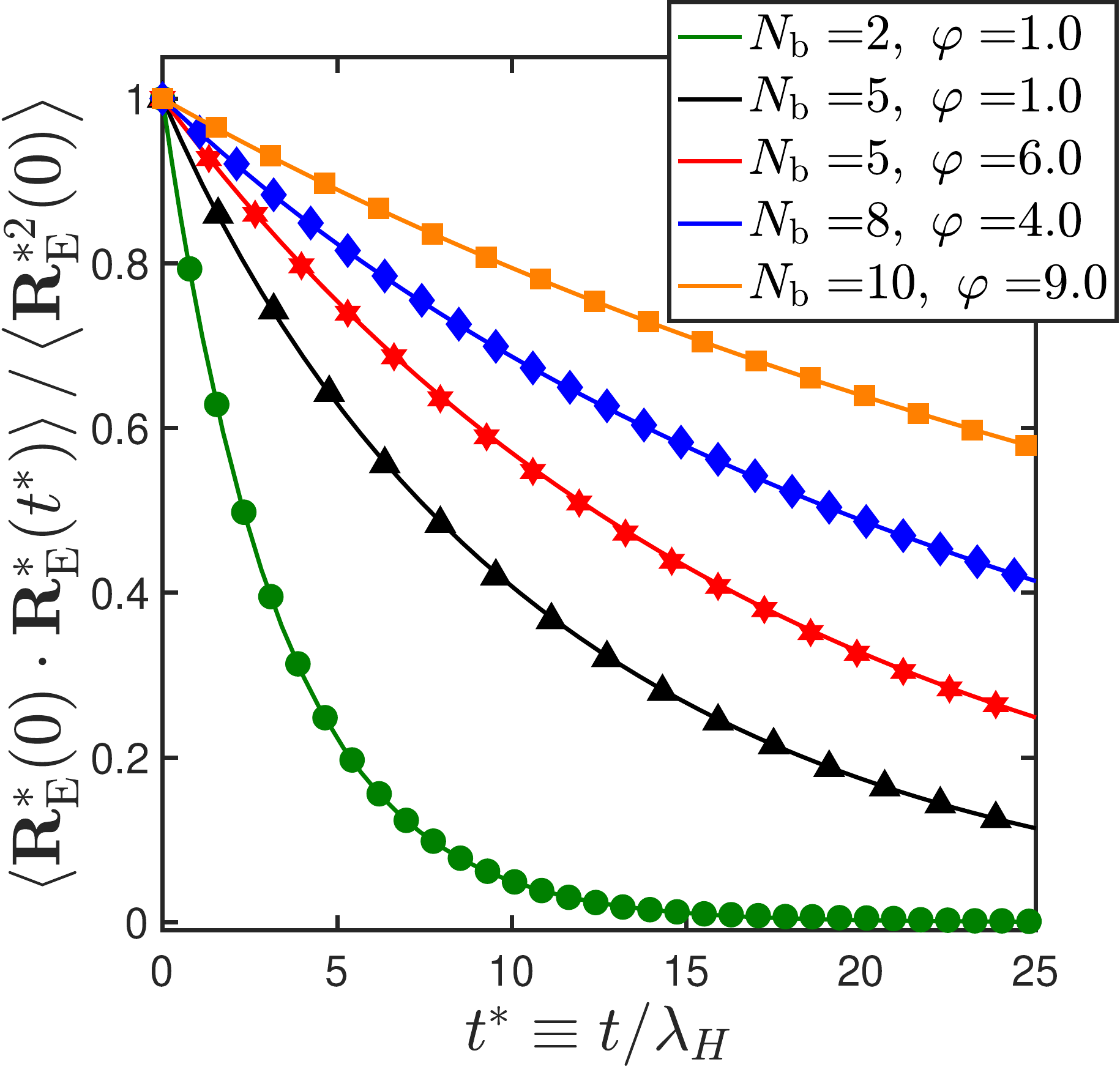}&
\includegraphics[width=2.7in,height=!]{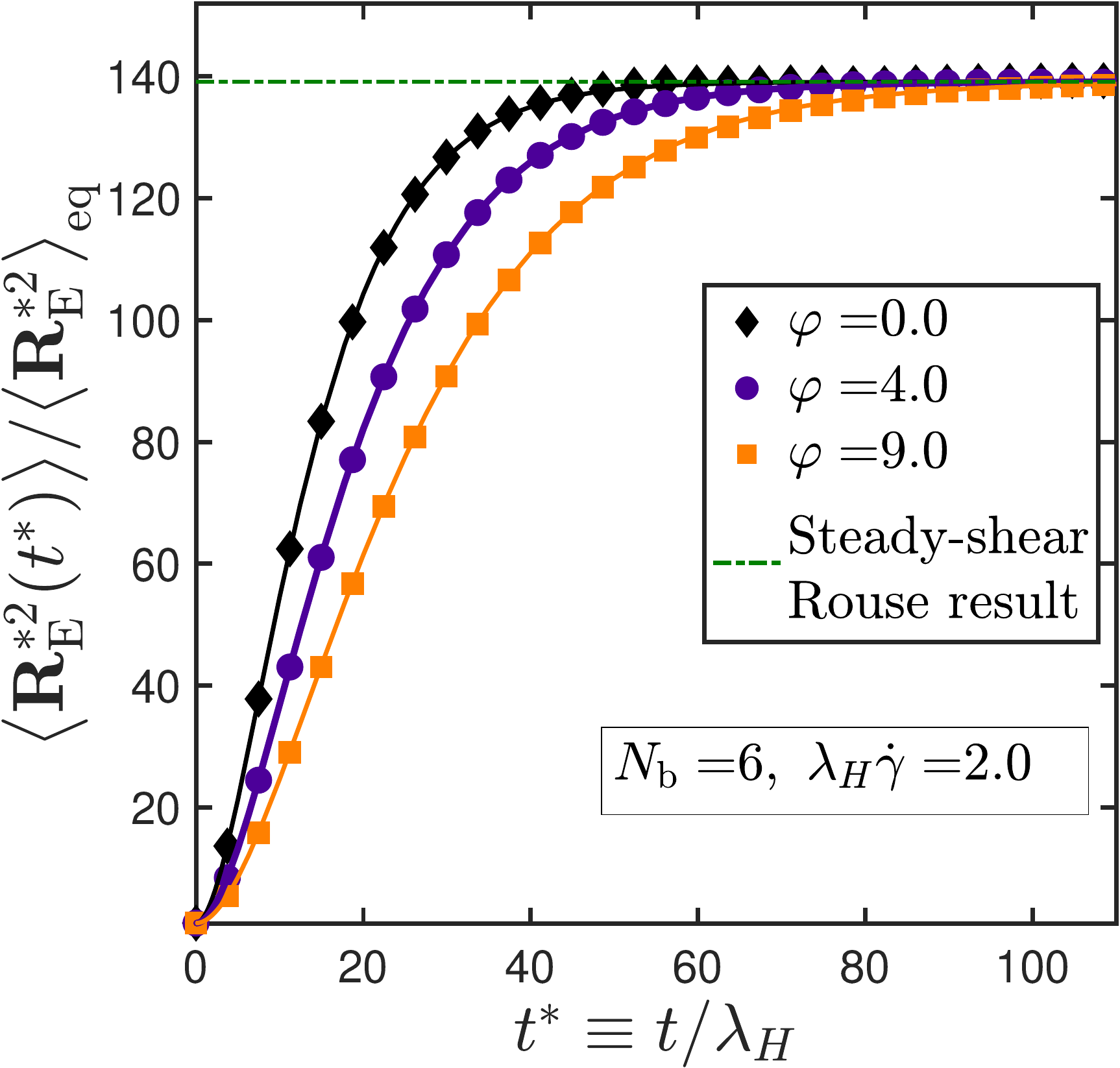}\\
(a)&(b)
\end{tabular}
\end{center}
\caption{\small BD simulation results of the preaveraged internal friction model (dispalyed as symbols) for: (a) Normalized autocorrelation of the end-to-end vector at equilibrium, and (b) transient evolution of the mean-squared end-to-end vector in shear flow, obtained by numerically integrating  the stochastic differential equation given by Eq.~(\ref{eq:sde_individual_dimless}) in both cases. The lines in (a) and (b) represent discrete RIF results given by Eq.~(\ref{eq:disc_exp_nm_dimless_main}) and Eq.~(\ref{eq:nmlzd_trans_re_t}), respectively. Error bars, which represent standard error of the mean, are roughly of the same size or smaller than the symbols used.}
\label{fig:equivalence_figs}
\end{figure*}

\section{\label{sec:rouse_preav}Equivalence between the bead-spring-dashpot chain with preaveraged internal friction and the discrete RIF model}

We show next that a Rouse model with preaveraged internal friction, constructed using the principles of PKT~\cite{Bird1987b,Ottinger1996}, is formally identical to the RIF model. A Hookean bead-spring-dashpot chain is considered, similar to the RIF model, except that the dashpot coefficient is taken to be $K$. The total force on a bead $\mu$ (not at the chain ends) due to internal friction is given by~\cite{ravibook}
\begin{equation}\label{eq:iv_force_form_bead}
\begin{split}
\bm{F}^{(\text{IV})}_{\mu}&=K\dashuline{\left(\dfrac{\bm{Q}_{\mu}\bm{Q}_{\mu}}{Q^2_{\mu}}\right)}\cdot\llbracket\dot{\bm{r}}_{\mu+1}-\dot{\bm{r}}_{\mu}\rrbracket\\
&-K\dashuline{\left(\dfrac{\bm{Q}_{\mu-1}\bm{Q}_{\mu-1}}{Q^2_{\mu-1}}\right)}\cdot\llbracket\dot{\bm{r}}_{\mu}-\dot{\bm{r}}_{\mu-1}\rrbracket
\end{split}
\end{equation}
where $\bm{Q}_{\mu}\equiv\bm{r}_{\mu+1}-\bm{r}_{\mu}$ is the connector vector joining the $\mu^{\text{th}}$ and the $(\mu+1)^{\text{th}}$ bead, and $\llbracket\cdots\rrbracket$ represents an average over the distribution of velocities in phase space. The equilibrium configurational distribution function for the model is unaltered by the presence of internal friction, and is simply given by the Gaussian distribution function for a Rouse chain. The preaveraging approximation entails a replacement of the underlined projection operators in Eq.~(\ref{eq:iv_force_form_bead}) by their average taken with respect to the equilibrium distribution function, which may be evaluated to be $\left(\boldsymbol{\delta}/3\right)$~\cite{doi-edwards}. The resultant internal friction force is then 
\begin{equation}
\bm{F}^{(\text{IV),preav.}}_{\mu}=\left(\dfrac{K}{3}\right)\llbracket\dot{\bm{r}}_{\mu+1}-2\dot{\bm{r}}_{\mu}-\dot{\bm{r}}_{\mu-1}\rrbracket
\end{equation}
which is identical to the RIF description of the same force as given by Eq.~(\ref{eq:iv_force_rif}). The governing stochastic differential equation for the preaveraged IV model has been derived in Appendix~\ref{sec:app_b}, and is numerically integrated using BD simulations. Notably, the use of the preaveraging approximation as a way to make flexible polymer models with internal friction analytically tractable was also suggested by Fixman~\cite{Fixman1988} several decades ago.

In Fig.~\ref{fig:equivalence_figs}, simulation results at equilibrium and in shear flow, obtained by numerically integrating the stochastic differential equation for the preaveraged IV model using BD simulations (shown as symbols) have been compared against discrete RIF predictions (indicated by lines), for several parameter values. The excellent agreement between the two model predictions establishes their equivalence. Furthermore, the preaveraged treatment predicts that internal friction only affects the time-evolution of the mean-squared end-to-end vector in shear flow, but not its steady-state value which is identical to the standard Rouse model prediction. 

\section{\label{sec:fluc_eqb}Effect of fluctuations at equilibrium}

\begin{figure*}[t]
\begin{center}
\begin{tabular}{c c}
\includegraphics[width=2.7in,height=!]{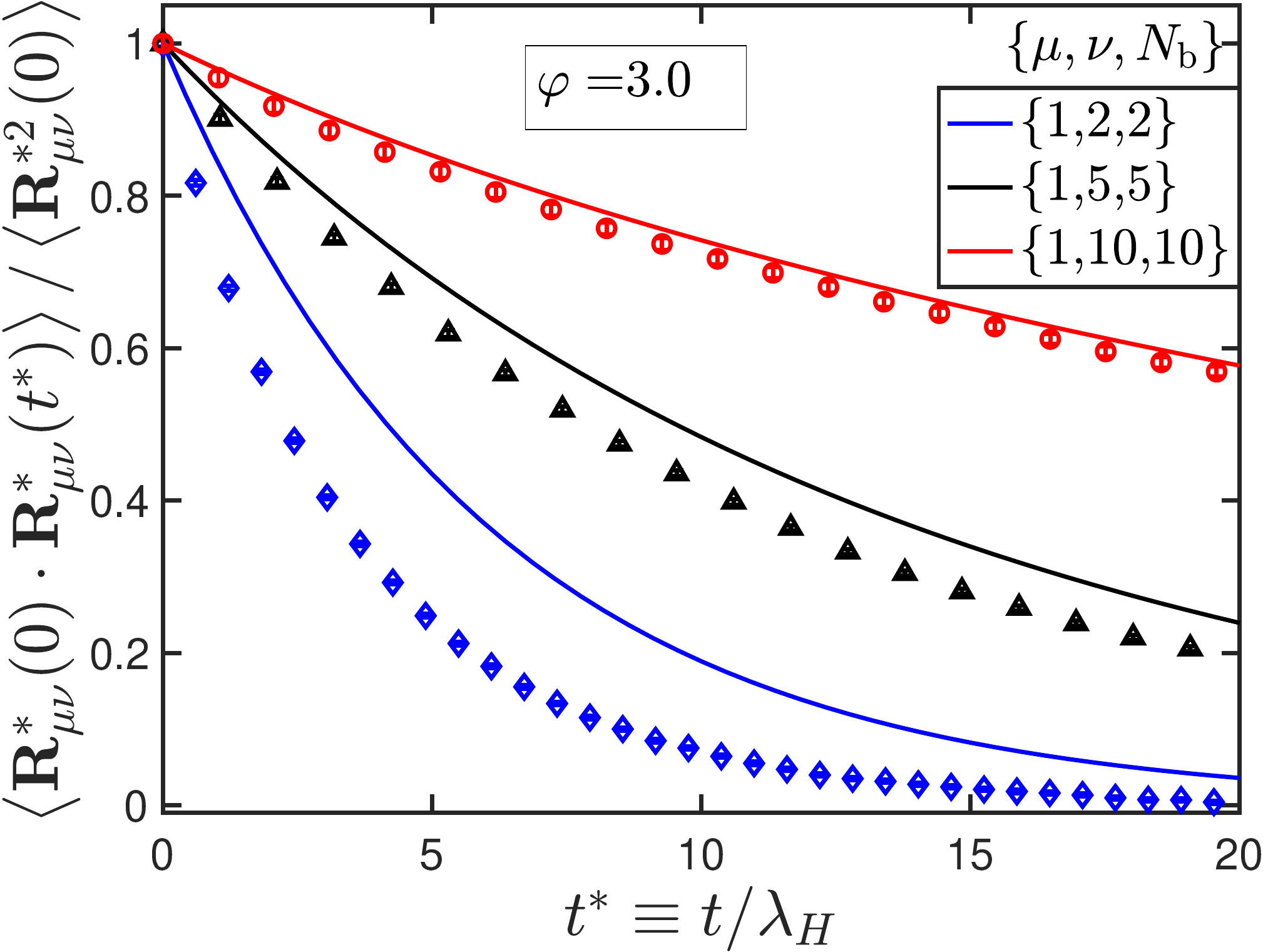}&
\includegraphics[width=2.7in,height=!]{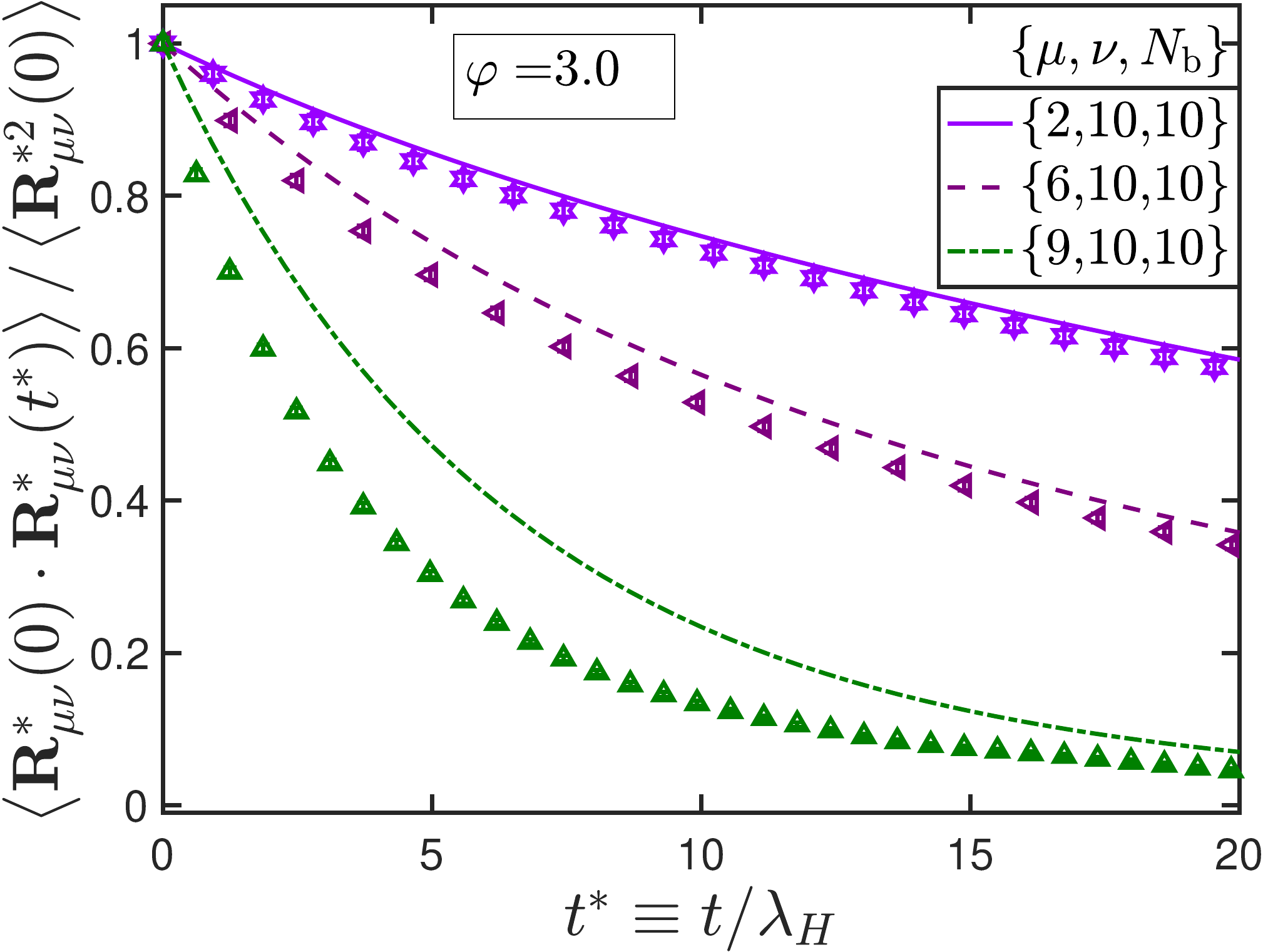}\\
(a) & (b)\\
\includegraphics[width=2.7in,height=!]{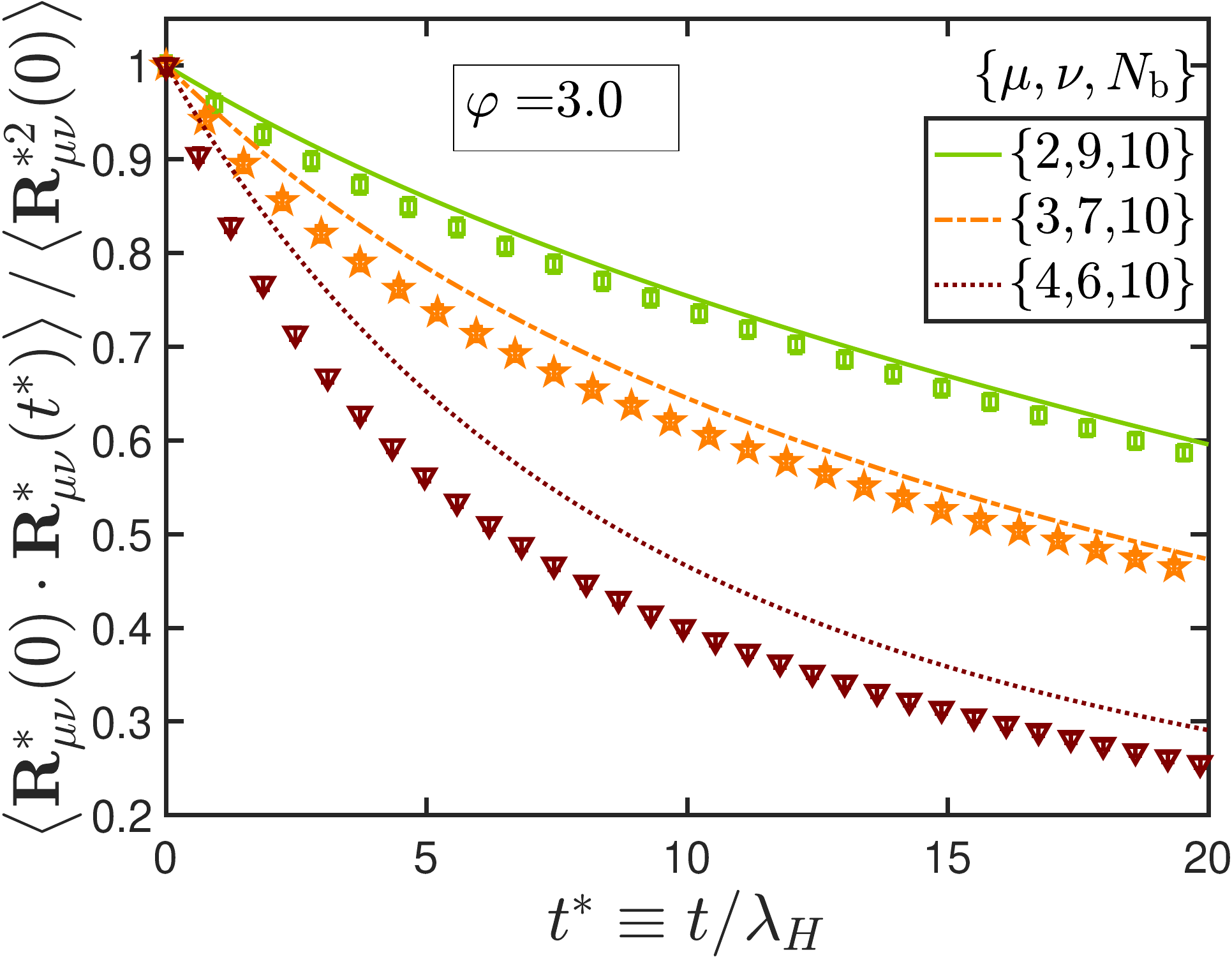}&
\includegraphics[width=2.7in,height=!]{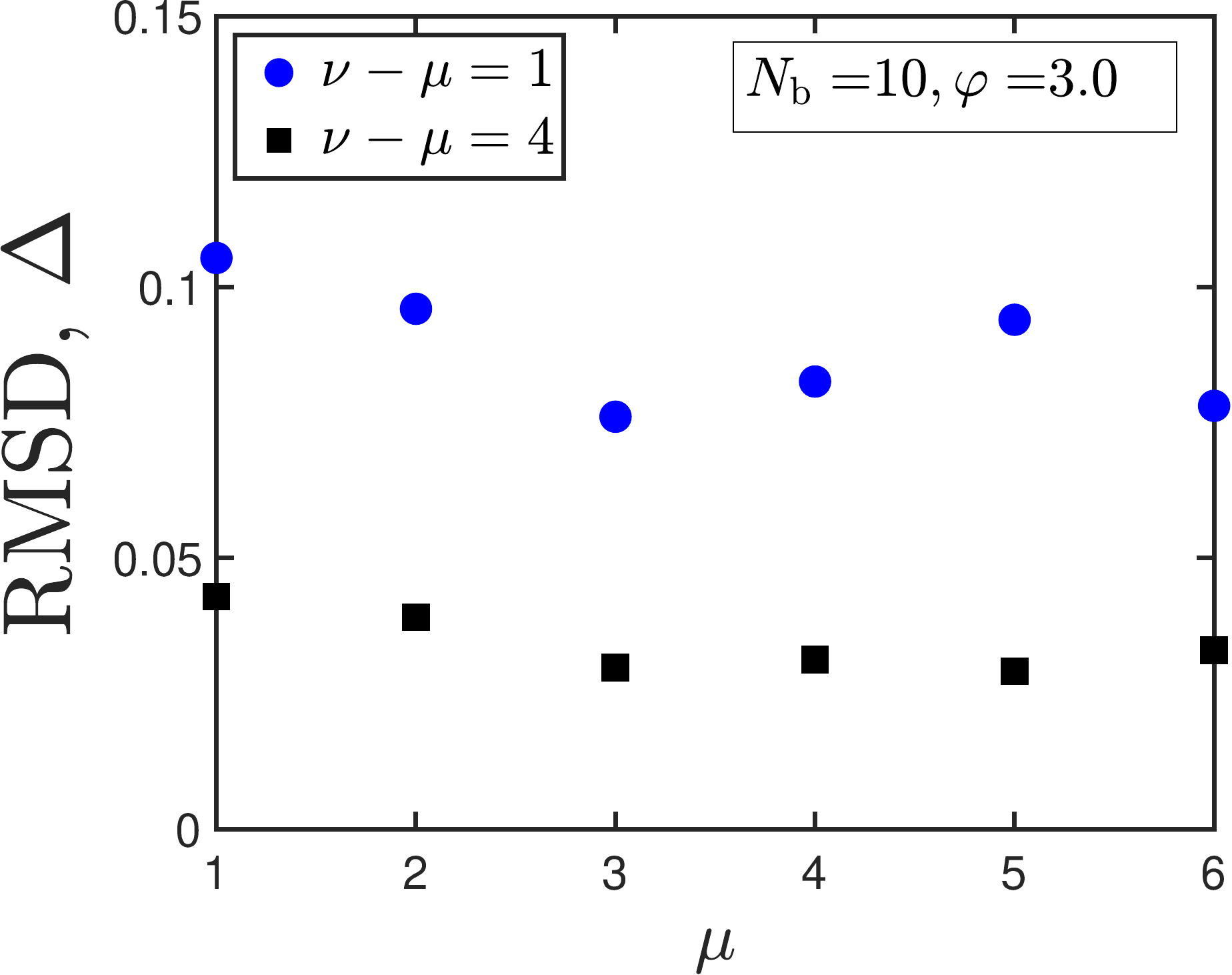}\\
(c) & (d)
\end{tabular}
\end{center}
\caption{\small Normalized autocorrelation of $\bm{R}^{*}_{\mu\nu}$, the vector connecting the $\mu^{\text{th}}$ and $\nu^{\text{th}}$ bead in a chain with $N_{\text{b}}$ beads, for (a) end-to-end, (b) end-to-interior, and (c) interior-to-interior cases. The notation $\{\mu,\nu,N_{\text{b}}\}$ is used to completely specify the vector originating at the bead index $\mu$ and terminating at $\nu$ in a chain with $N_{\text{b}}$ beads. Lines represent preaveraged model results given by Eq.~(\ref{eq:mu_nu_correl_in_paper}). Symbols are BD simulation results of the Rouse model with fluctuating internal friction. An internal friction parameter of $\varphi=3.0$ has been used for all the cases.  Subfigure (d) represents the root-mean-squared difference between the two model results for various values of the interbead separation $\nu-\mu$, quantified as 
$\Delta=\sqrt{\dfrac{1}{N_{\text{points}}}\sum_{i=1}^{N_{\text{points}}}\left[y^{\text{(BD)}}_{i}-y^{\text{(analytical)}}_{i}\right]^2}$, where $y^{\text{(BD/analytical)}}_{i}$ refers to the simulation/analytical result at the $i^{\text{th}}$ datapoint, with $N_{\text{points}}$ denoting the number of time-instances at which the simulation results have been obtained.}
\label{fig:autocorr}
\end{figure*}

As the next step, the accuracy of the preaveraging approximation is compared against the exact numerical solution we have derived recently~\cite{kailasham2021rouse}, in which the original non-preaveraged form of the internal friction force given by Eq.~(\ref{eq:iv_force_form_bead}) is used. 

In Fig.~\ref{fig:autocorr}, the normalized autocorrelation of  $\bm{R}^{*}_{\mu\nu}$ in a chain with $N_{\text{b}}$ beads predicted by the discrete RIF model (indicated by lines) is compared against exact BD simulation results of a model with fluctuating internal friction (shown as symbols). The notation $\{\mu,\nu,N_{\text{b}}\}$ uniquely identifies the vector originating at the bead index $\mu$ and terminating at $\nu$ in a chain of $N_{\text{b}}$ beads. The parameter space specified by $\{\mu,\nu,N_{\text{b}}\}$ may broadly be classified into three topological classes~\cite{DesCloizeaux1980,Toan2008,Samanta2014,kumari2021spatiotemporal}, and these categories have been considered in Fig.~\ref{fig:autocorr}~(a)-(c). In Fig.~\ref{fig:autocorr}~(a), the end-to-end case is considered, where $\mu$ and $\nu$ are taken to be the terminal beads in a chain.  Fig.~\ref{fig:autocorr}~(b) represents the end-to-interior topology class, where $\nu$ is taken to be a terminal bead, and $\mu$ is chosen from the interior set of beads, while Fig.~\ref{fig:autocorr}~(c) represents the interior-to-interior case wherein both $\mu$ and $\nu$ are taken to be interior beads. In all the three cases, the difference between the fluctuating IV model and the preaveraged one is seen to diminish with an increase in $\nu-\mu$, i.e., the number of beads between the $\mu$ and $\nu$ positions. A similar qualitative trend was observed for other values of $\varphi$, and consequently, $\varphi=3.0$, has been used for all the cases considered in Fig.~\ref{fig:autocorr}. In Fig.~\ref{fig:autocorr}~(d), the root-mean-squared difference (RMSD, $\Delta$) between the two model results are plotted against the bead number $\mu$, at two different values of the interbead separation, $\nu-\mu$. It is observed that the RMSD decreases with an increase in the interbead separation, and is fairly insensitive to the specific choice of the bead number $\mu$. This implies that the end-to-interior, and the interior-to-interior topology classes may be considered equivalent with regards to the measured deviation between the preaveraged and fluctuating IV results.

There are two implications to the trend observed in Fig.~\ref{fig:autocorr}: firstly, the preaveraged model is sufficiently accurate for characterizing the bead connector vector correlations of longer segments, as compared to local correlations corresponding to shorter chain segments, and secondly, the preaveraged model may satisfactorily be used for end-to-end vector reconfiguration time calculations provided that a sufficiently fine enough level of discretization (i.e., large enough $N_{\text{b}}$) is chosen. Note, however, that the choice of $N_{\text{b}}$ is not arbitrary, and its largest permitted value is the number of Kuhn segments in the underlying polymer molecule. For example, the RIF model~\cite{Samanta2016165,Soranno2017,Soranno2018} has been used for studying internal friction effects in the sixty-seven amino-acid residue cold shock protein at various concentrations of the denaturant, guanindinium chloride (GdmCl). The protein has a Kuhn segment length of about five residues at 6M GdmCl, which implies that the number of Kuhn segments in the molecule, and consequently the finest level of discretization, $N_{\text{b}}$, is $(67/5)\approx13$. From Fig.~\ref{fig:autocorr}, it would appear that at such values of $N_{\text{b}}$, the use of the preaveraged IV model for the calculation of the reconfiguration time would be justified for an internal friction parameter of $\varphi=3.0$. The discrepancy between the two model predictions, however, is expected to increase with the internal friction parameter. Furthermore, internal friction has also been observed in a synthetic tryptophan cage molecule~\cite{Qiu2002,Qiu2004} with twenty-residues, whose Kuhn segment length has not been reported, and it would appear that fluctuations in internal friction should be included for modeling such small molecules, given the constraint on the level of discretization.

\begin{figure}[t]
\centering
\includegraphics[width=3.1in,height=!]{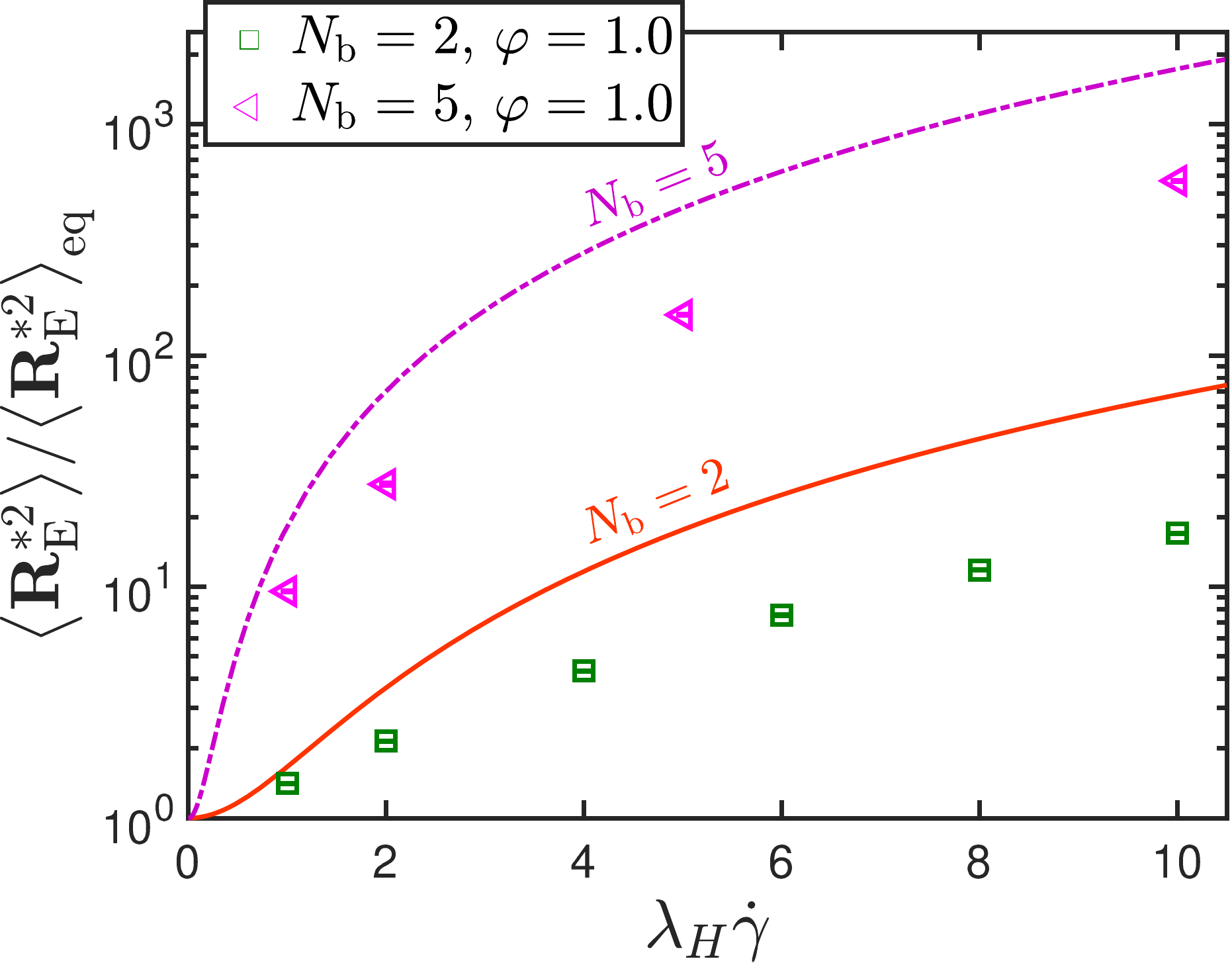}  
\caption{\small Normalized steady-state mean-squared distance as a function of dimensionless shear rate for chains in shear flow. Lines correspond to preaveraged model results given by Eq.~(\ref{eq:nmlzd_steady_re_t}).  Symbols are BD simulation results on the Rouse model with fluctuating internal friction.}
\label{fig:shear_compare}
\end{figure}

\begin{figure}[pthb]
\begin{center}
\begin{tabular}{c}
\includegraphics[width=8cm,height=!]{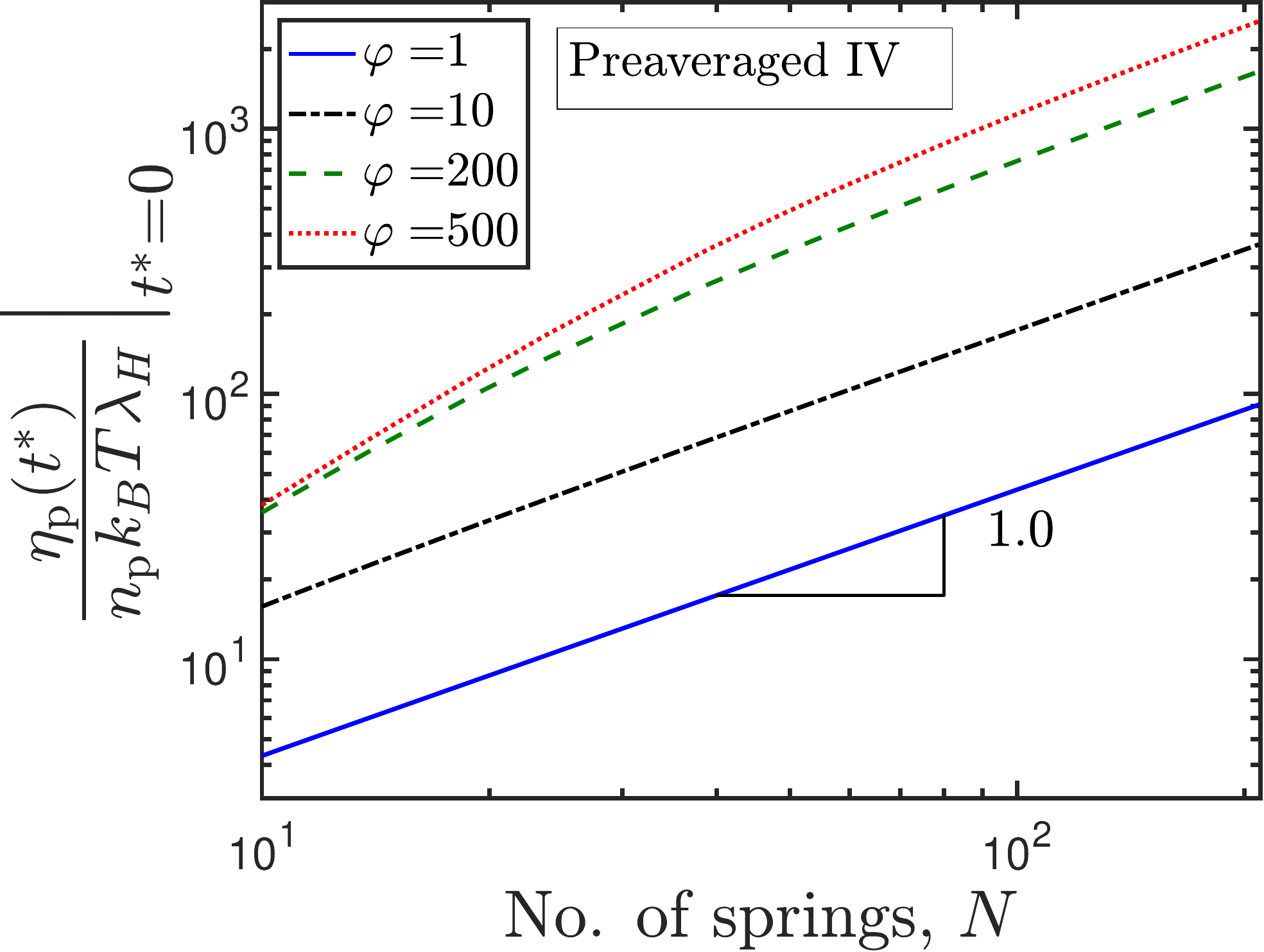}\\
(a)\\[10pt]
\includegraphics[width=8cm,height=!]{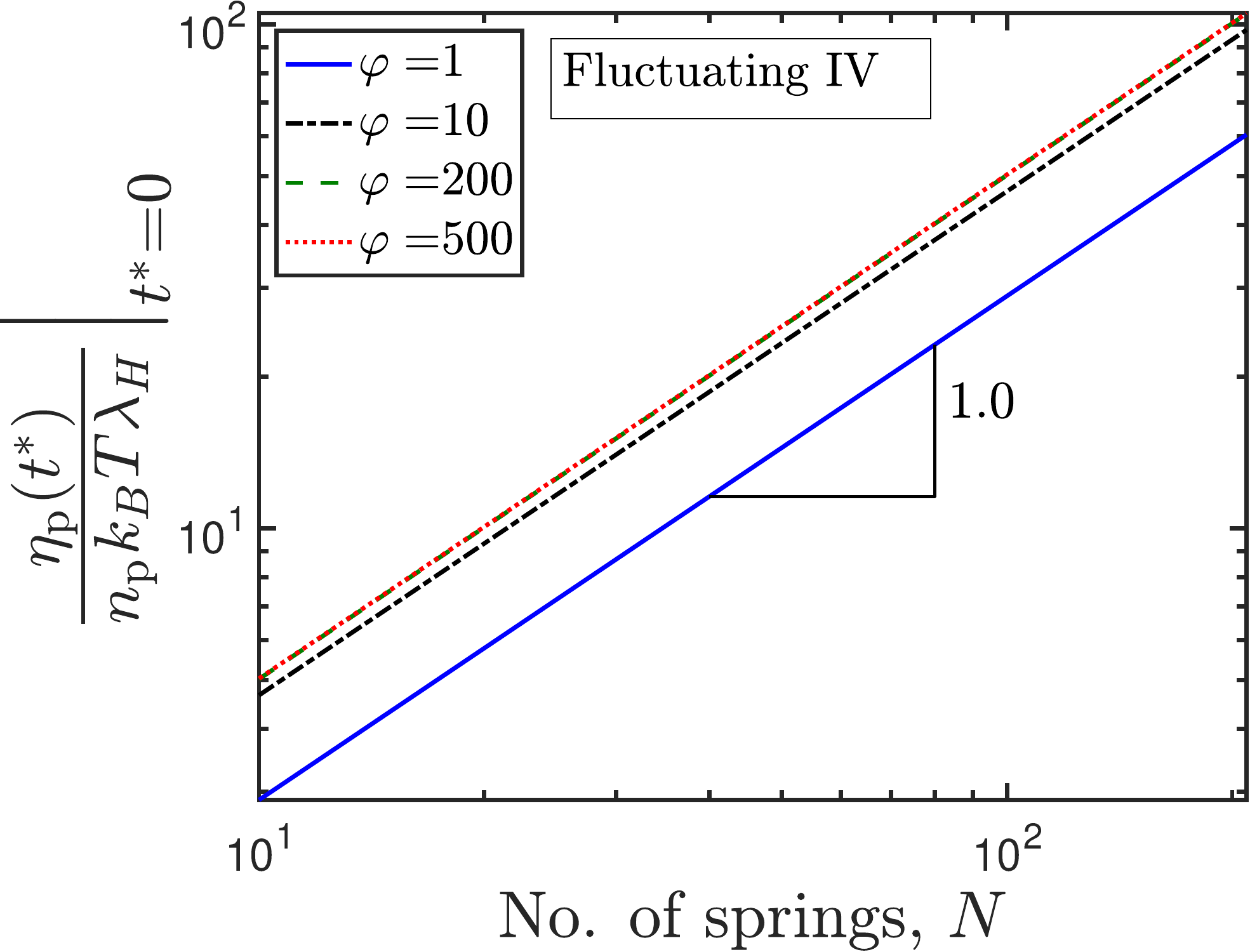}\\
(b)\\[10pt]
\includegraphics[width=8.1cm,height=!]{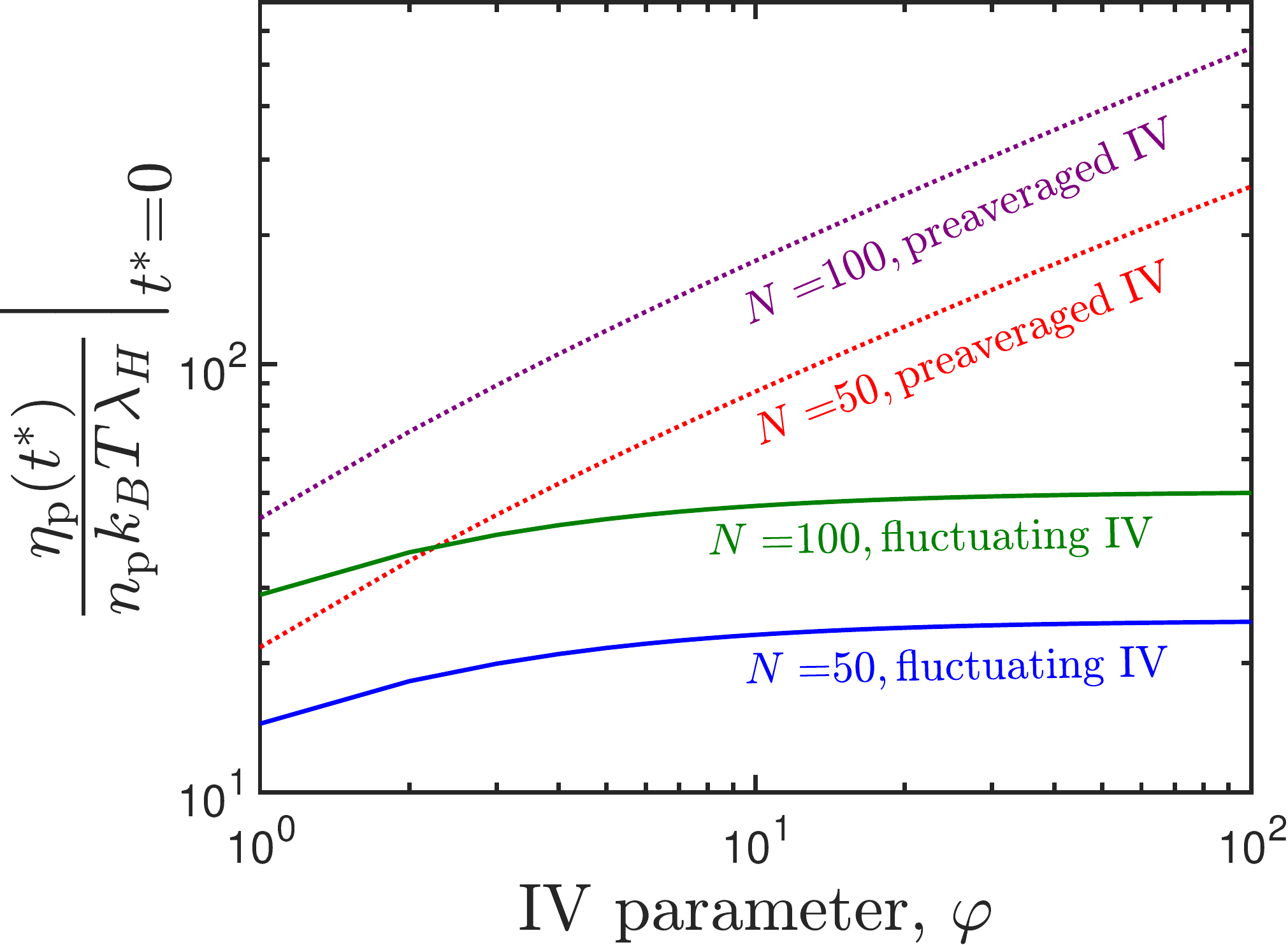}\\
(c)\\[10pt]
\end{tabular}
\end{center}
\caption{\small Stress jump as a function of chain length for (a) preaveraged IV and (b) fluctuating IV models. In (c) the stress jump is plotted as a function of the internal friction parameter, for two different chain lengths. Preaveraged model results are given by Eq.~(\ref{eq:preav_iv_eta_jump}), while the fluctuating IV predictions are obtained using the semi-analytical approximation given by Manke and Williams~\cite{Manke1988}.}
\label{fig:sjump_scaling}
\end{figure}

\section{\label{sec:fluc_shear_flow}Effect of fluctuations in shear flow}

In Fig.~\ref{fig:shear_compare}, the normalized, steady-state mean-squared end-to-end distance of a chain in simple shear flow is plotted as a function of the dimensionless shear rate. The lines, which represent the preaveraged model results, coincide with the simple Rouse model predictions implying that the steady-shear values are unaffected by the internal friction parameter, as also evident from Fig.~(\ref{fig:equivalence_figs})~(b). The model with fluctuating internal friction, however, predicts that the extension in shear-flow is a function of the internal friction parameter.

\begin{figure}[t]
\begin{center}
\begin{tabular}{c}
\includegraphics[width=3.4in,height=!]{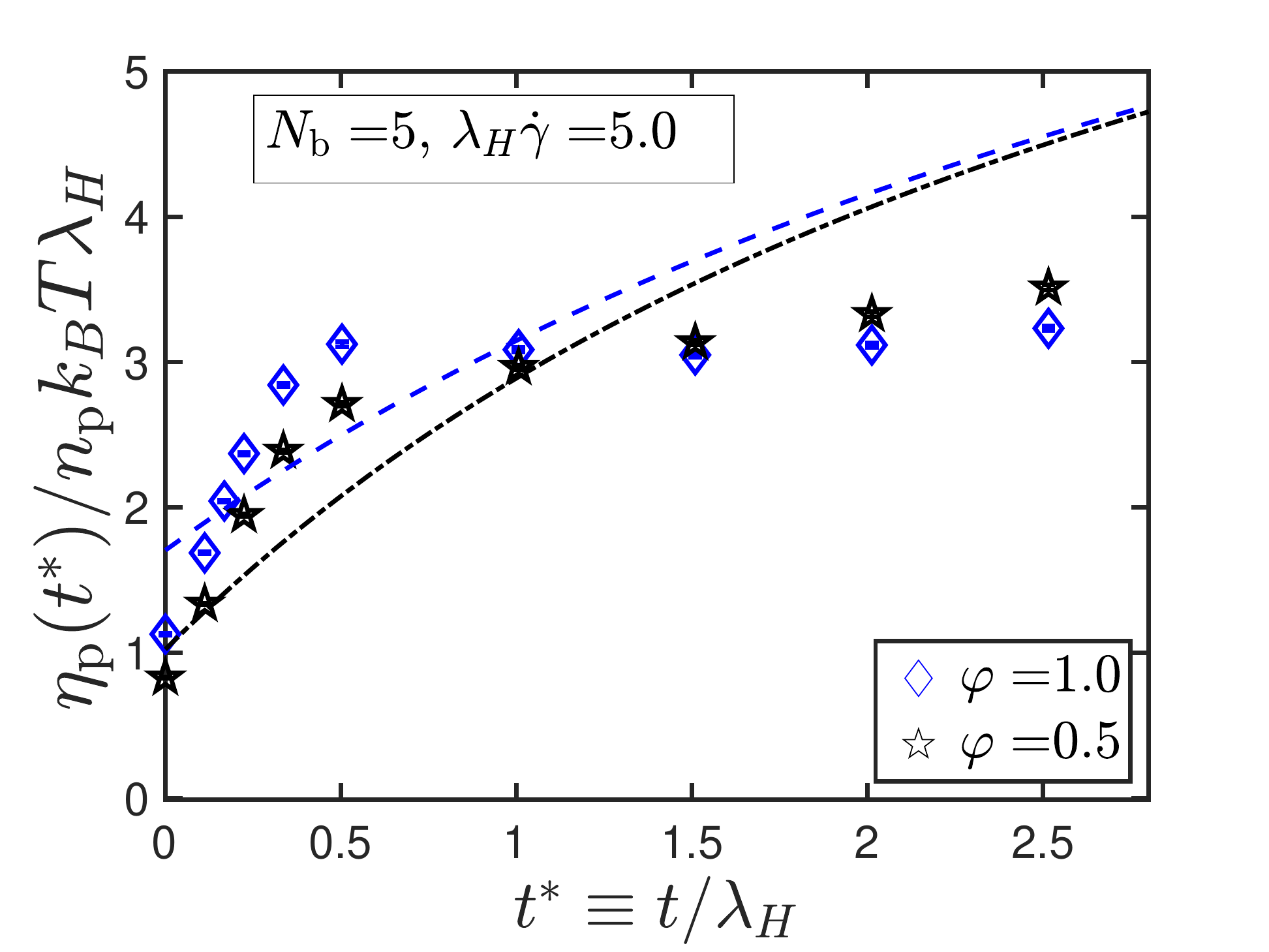}\\
(a)\\
\includegraphics[width=3.4in,height=!]{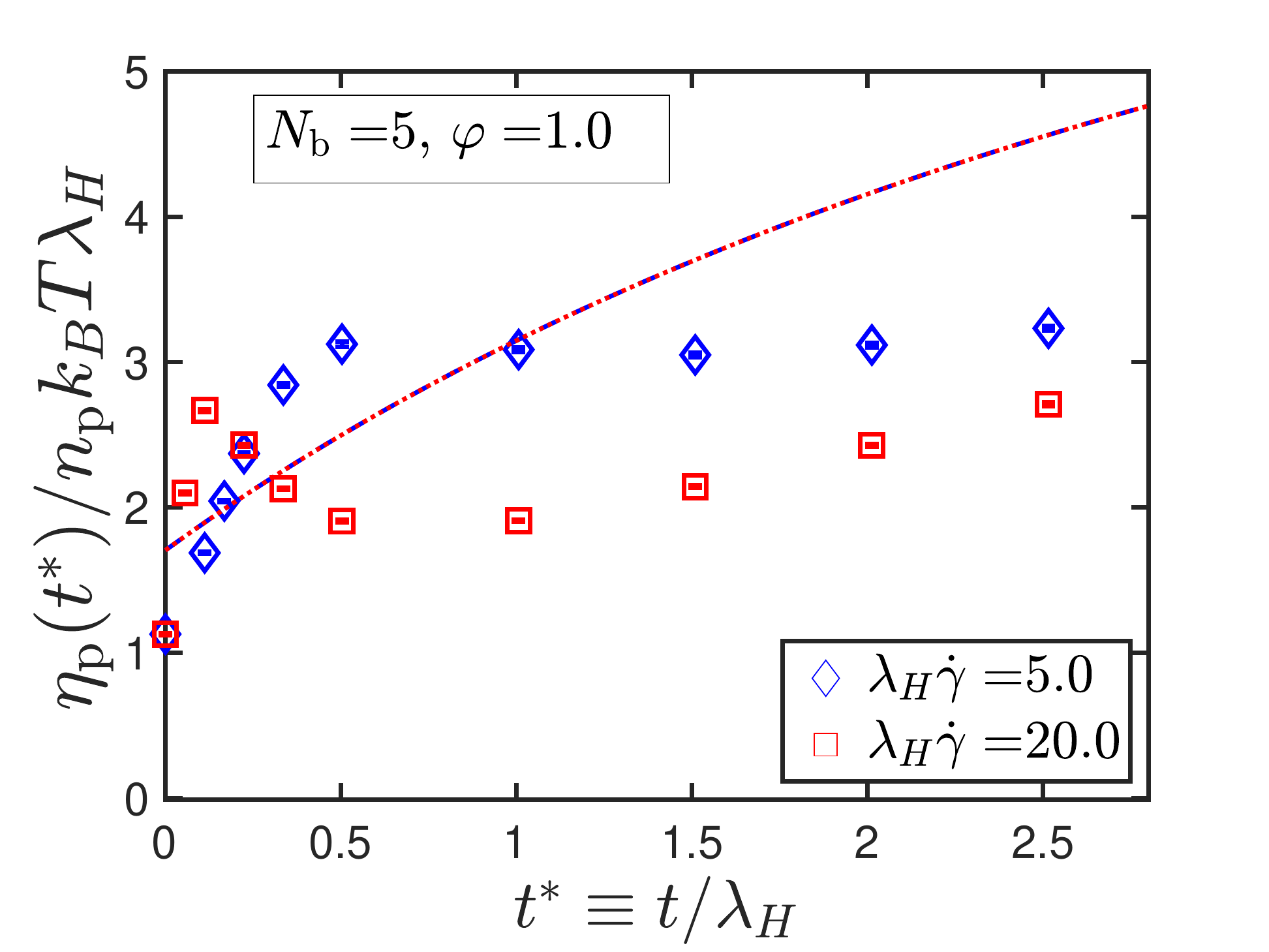}\\
(b)
\end{tabular}
\end{center}
\vspace{-5pt}
\caption{\small The effect of (a) internal friction parameter, and (b) shear rate on the time evolution of the dimensionless shear viscosity for a five-bead Rouse chain with internal friction. Lines correspond to preaveraged model predictions given by Eq.~(\ref{eq:preav_iv_trans_eta}). Symbols are BD simulation results of the Rouse model with fluctuating internal friction.}
\label{fig:shear_compare_fluc}
\end{figure}

There exist no prior studies of the viscometric functions predicted by the Rouse model with preaveraged internal friction. We have derived (as detailed in Appendix~\ref{sec:derv_stress_tensor}) an expression for the stress tensor by using the Giesekus formula~\cite{Bird1987b}, since this choice has been shown to lead to thermodynamically consistent results for models with fluctuating IV~\cite{Schieber1994}. An expression for the transient evolution of shear viscosity may be derived to be
\begin{equation}\label{eq:preav_iv_trans_eta}
\begin{split}
\dfrac{\eta_{\text{p}}(t^*)}{n_{\text{p}}k_BT\lambda_{H}}&=\left(\dfrac{3}{3+2\varphi}\right)\sum^{N}_{m,n,q=1}\Pi_{mq}\mathcal{L}_{mn}\Pi_{nq}I_{q}(t^*)\\[5pt]
&+{2\,\text{tr}\left[\bm{\mathcal{C}}-\dfrac{3}{3+2\varphi}\bm{\mathcal{S}}\right]};
\end{split}
\end{equation}
with
\begin{equation}\label{eq:i_def_trans}
I_{q}(t^*)=2\left(\dfrac{\widetilde{b}_q}{\widetilde{a}_q}\right)\left\{1-\exp\left[-\left(\dfrac{\widetilde{a}_q}{1+2\theta}\right)\dfrac{t^{*}}{2}\right]\right\}
\end{equation}
The detailed steps for the derivation of Eqs.~(\ref{eq:preav_iv_trans_eta}) and~(\ref{eq:i_def_trans}), along with the definitions of the quantities that appear in these equations, have been provided in Appendix~\ref{sec:derv_stress_tensor}. The validity of Eq.~\ref{eq:preav_iv_trans_eta} has been established by comparison with BD simulations of the preaveraged IV model, as illustrated in Fig.~\ref{fig:shear_compare2} in Appendix~\ref{sec:derv_stress_tensor}.

An important rheological consequence of internal friction is the appearance of a discontinuous, shear-rate-independent, jump in viscosity at the inception of flow~\cite{Manke1988,Hua1995,Kailasham2018}. This phenomenon, called ``stress jump" is not predicted by other bead-spring-chain models. From Eq.~\ref{eq:i_def_trans}, it is clear that the function $I_{q}(t^*)\to0$ as $t^{*}\to 0$, and the stress jump for the preaveraged IV model is therefore given by 
\begin{equation}\label{eq:preav_iv_eta_jump}
\begin{split}
\left. \frac{\eta_{\text{p}}(t^{*})}{n_{\text{p}}k_BT\lambda_{H}}\right\vert_{t^{*}=0}&={2\,\text{tr}\left[\bm{\mathcal{C}}-\dfrac{3}{3+2\varphi}\bm{\mathcal{S}}\right]}
\end{split}
\end{equation}
In Fig.~\ref{fig:sjump_scaling}, the stress jumps predicted by the preaveraged and the fluctuating IV models are plotted as a function of the chain lengths and the internal friction parameter. A semi-analytical approximation for the stress jump in Rouse chains with fluctuating IV has been derived by Manke and Williams~\cite{{Manke1988}}, and compares excellently against exact BD simulation results~\cite{kailasham2021rouse}, with the accuracy of the approximation observed to improve with an increase in the number of beads. This approximate solution has therefore been used to plot Fig.~\ref{fig:sjump_scaling}~(b) due to the computational intensity of performing BD simulations for large chain lengths~\cite{kailasham2021rouse}. From Figs.~\ref{fig:sjump_scaling}~(a) and ~\ref{fig:sjump_scaling}~(b), it is observed that while the fluctuating IV model predicts that the stress jump scales linearly with the chain length for values of the internal friction parameter spanning two orders of magnitude, a similar linear dependence in the preaveraged model predictions is pushed to larger values of the number of springs, $N$, as the internal friction parameter is increased. From Fig.\ref{fig:sjump_scaling}~(c), it is observed that for a given value of the internal friction parameter, $\varphi$, and chain length, the stress jump predicted by the fluctuating IV model is lower than that predicted by the preaveraged model. Furthermore, for a given value of the chain length, the stress jump predicted by the fluctuating IV model saturates with an increase in the internal friction parameter. No such saturation, however, is predicted by the preaveraged IV model. 

In Fig.~\ref{fig:shear_compare_fluc}, the transient evolution of shear-viscosity for models with preaveraged and fluctuating IV are compared for a five-bead chain. As seen from Fig.~\ref{fig:shear_compare_fluc}~(a), at a fixed value of the shear rate, there is a qualitative difference in the transient evolution of viscosity predicted by the two models, for two different values of the internal friction parameter. While the preaveraged model prediction for the viscosity grows smoothly with time, the fluctuating IV results display a sharper initial rise followed by a more gradual approach to their steady-state values. In Fig.~\ref{fig:shear_compare_fluc}~(b), the transient response is plotted for a fixed value of the internal friction parameter, at two different shear rates. The preaveraged model prediction is independent of the shear rate, and grows monotonically, while the viscosity predicted by the model with fluctuations is shear-rate-dependent, going through a local maximum for larger shear rates, as seen clearly for the $\lambda_{H}\dot{\gamma}=20.0$ case. In our previous work~\cite{Kailasham2018}, the transient shear viscosity of dumbbells with IV was shown to exhibit an overshoot at high shear rates, with the viscosity rising above its steady-state value before settling to its asymptotic limit.

Furthermore, in the long-time limit, the preaveraged IV model predicts a shear-rate independent, constant value of the viscosity, equal to the Rouse viscosity for a given chain length, independent of the internal friction parameter. Our recent work~\cite{kailasham2021rouse} shows that Rouse chains with fluctuating IV, however, exhibit a shear-thinning followed by a shear-thickening of the steady-state viscosity with respect to the shear rate, with the internal friction parameter governing the onset and extent of the observed shear-thickening. 

\section{\label{sec:concl_fluc_imp}Conclusions}

The results of this paper clearly indicate that fluctuations in internal friction significantly affect the dynamics of polymer molecules away from equilibrium. While a majority of experiments and simulations~\cite{Wensley2010,Soranno201217800,ja211494h,Cheng2013,Ameseder2018} over the last two decades on understanding the effects of internal friction on biomolecule dynamics have focused on equilibrium measurements, such as reconfiguration and folding times, the effect of this phenomenon on the probability distribution of polymer extensions in coil-stretch transitions during turbulent flow has recently garnered attention~\cite{Vincenzi2020}. We anticipate that the present work will provide a theoretical framework for discerning the effects of internal friction in out-of-equilibrium systems. 

Quantitative comparisons against experiments would require the incorporation of hydrodynamic interaction effects~\cite{Sasmal2016,Prakash2019}. However, the solution of coarse-grained models which account for fluctuations in both internal friction and hydrodynamic interactions (with more than two beads~\cite{Kailasham2018}), have additional challenges~\cite{kailasham2021rouse} that have not been addressed so far.

\begin{acknowledgements}
Numerical simulations were performed on the MonARCH and MASSIVE computer clusters of Monash University, and the SpaceTime-2 computational facility of IIT Bombay. R. C. acknowledges SERB for funding (Project No. MTR/2020/000230 under MATRICS scheme). We also acknowledge the funding and general support received from the IITB-Monash Research Academy.
\end{acknowledgements}

\vspace{-20pt}

\appendix

\section{\label{sec:d_rif} Discrete version of Rouse model with internal friction}

\subsection{\label{sec:app_a}Analytical solution to the discrete RIF model}

The Rouse matrix in Eq.~(\ref{eq:bead_govern_with_shear}) is of size $N_{\text{b}}\times\,N_{\text{b}}$ and has the following structure~\cite{Verdier1966}
\begin{equation}\label{eq:rouse_mat_def}
\bm{A}^{\text{(R)}} = 
\begin{pmatrix}
1& -1 & {0} & \cdots & {} & {0}\\
-1 & 2& -1 & {0} & \cdots & {0} \\
{0} & -1 & 2 & -1 &\cdots & {} \\
\vdots  & \vdots & \vdots & {}& {} & {} \\
{0} & {0}& \cdots & {-1} & 2 & -1 \\
{0} & {0}& \cdots & {0} & -1 & 1
\end{pmatrix}
\end{equation}
The elements of the orthogonal matrix $\bm{\Omega}$ which project the bead-positions into normal-mode space are given by~\cite{Verdier1966,Kopf1997}
\begin{equation}\label{eq:ortho_mat_def}
{\Omega}_{\mu n}=\left(\dfrac{2-\delta_{n0}}{N_{\text{b}}}\right)^{1/2}\cos\left[\left(\mu-\dfrac{1}{2}\right)\dfrac{n\pi}{N_{\text{b}}}\right]
\end{equation}
where $\mu=1,2,3,...N_{\text{b}}$ and $n=0,1,2,...(N_{\text{b}}-1)$. The columns of $\boldsymbol{\Omega}$ are eigenvectors of $\bm{A}^{\text{(R)}}$, which means
\begin{equation}\label{eq:ortho_prop1}
\sum_{\mu}\Omega_{\mu m}\Omega_{\mu n} = \delta_{mn}
\end{equation}
\begin{equation}\label{eq:ortho_prop2}
\sum_{n}\Omega_{\mu n}\Omega_{\nu n} = \delta_{\mu\nu}
\end{equation}
\begin{equation}\label{eq:ortho_prop3}
\sum_{\mu}\sum_{\nu}\Omega_{\mu n}A^{\text{(R)}}_{\mu \nu}\Omega_{\nu m} = a_{m} \delta_{nm}
\end{equation}
where $a_m$ refers to the eigenvalues of $\bm{A}^{\text{(R)}}$, given by
\begin{equation}\label{eq:eigval_def}
a_{m}=4\sin^2\left(\dfrac{m\pi}{2N_{\text{b}}}\right)\,;\quad m=0,1,2,...,(N_{\text{b}}-1)
\end{equation}
Applying the transformation to normal co-ordinates, $\bm{X}_j(t)=\sum_{\mu}\Omega_{\mu j}\bm{r}_{\mu}(t)$, to Eq.~(\ref{eq:bead_govern_with_shear}), the governing equation in normal-mode coordinates becomes
\begin{widetext}
\begin{equation}\label{eq:mode_govern_with_shear}
\dfrac{d\bm{X}_p}{dt}=-\dfrac{Ha_p}{\zeta\left(1+\theta a_p\right)}\bm{X}_p+\left(\dfrac{1}{1+\theta a_p}\right)\boldsymbol{\kappa}\cdot\bm{X}_p+\bm{g}_p(t)
\end{equation}
where $\theta=\left(K/3\zeta\right)=\varphi/3$, and the moments of the noise vector, $\bm{g}_p(t)$, are given as follows
\begin{equation}\label{eq:norm_mode_noise_rel}
\left<{g}^{\alpha}_{p}\right>=0;\quad \left<{g}^{\alpha}_{p}(t){g}^{\beta}_{q}(t')\right>=\dfrac{2k_BT}{\zeta_p}\delta_{pq}\delta^{\alpha\beta}\delta(t-t')
\end{equation}
with $H_p=Ha_p$, and $\zeta_p=\zeta\left(1+\theta a_p\right)$. The indices $\alpha$ and $\beta$ in Eq.~(\ref{eq:norm_mode_noise_rel}) run from 1 to 3. With $\boldsymbol{\kappa}=\bm{0}$, Eq.~(\ref{eq:mode_govern_with_shear}) represents the equation of motion of a Brownian harmonic oscillator moving in a potential of stiffness $H_p$ and experiencing a friction coefficient $\zeta_p$. For such an oscillator, it is known that~\cite{Verdier1966,doi-edwards,Kopf1997}  
\begin{equation}\label{eq:correl_mode}
\left<\bm{X}_p(0)\cdot\bm{X}_q(t)\right>=\dfrac{3k_BT}{H_p}\delta_{pq}e^{-t/\tau_p},
\end{equation}
where 
\begin{equation}\label{eq:taup_def}
\tau_{p}=\zeta_{p}/H_{p}=\left(\zeta/4H\right)\sin^{-2}\left(p\pi/2N_{\text{b}}\right)+\left(K/3H\right),
\end{equation}
The expression for the bead position vectors in terms of the normal coordinates is given by
\begin{equation}
\bm{r}_{\mu}=\sum_{j=0}^{N_{\text{b}}-1}\Omega_{\mu\,j}\bm{X}_{j}(t)
\end{equation}
and the vector joining beads $\mu$ and $\nu$ is written as
\begin{equation}\label{eq:interbead_init}
\bm{R}_{\mu\nu}(t) \equiv\bm{r}_{\nu}(t)-\bm{r}_{\mu}(t)
=\sqrt{\dfrac{2}{N_{\text{b}}}}\sum_{q=1}^{N_{\text{b}}-1}\Biggl\{\cos\left[\left(\nu-\dfrac{1}{2}\right)\dfrac{q\pi}{N_{\text{b}}}\right]-\cos\left[\left(\mu-\dfrac{1}{2}\right)\dfrac{q\pi}{N_{\text{b}}}\right]\Biggr\}\bm{X}_{q}(t)
\end{equation}
Using Eqs.~(\ref{eq:correl_mode}) and (\ref{eq:interbead_init}), the normalized autocorrelation of the interbead connector vector is
\begin{equation}\label{eq:exp_nm_dim}
\begin{split}
\dfrac{\left<\bm{R}_{\mu\nu}(0)\cdot\bm{R}_{\mu\nu}(t)\right>}{\left<\bm{R}^{2}_{\mu\nu}(0)\right>}=\left[\dfrac{2}{N_{\mathrm{b}}|\nu-\mu|}\right]\,\Biggl[&\sum_{p=1}^{N_{\text{b}}-1}\left\{\cos\left[\left(\nu-\dfrac{1}{2}\right)\dfrac{p\pi}{N_{\mathrm{b}}}\right]-\cos\left[\left(\mu-\dfrac{1}{2}\right)\dfrac{p\pi}{N_{\mathrm{b}}}\right]\right\}^2\left(\dfrac{1}{a_p}\right)e^{-t/\tau_p}\Biggr]
\end{split}
\end{equation}
The choice of the noise term in normal mode space [Eq.~(\ref{eq:norm_mode_noise_rel})] ensures that the mean-squared value of the segmental vector $\bm{R}_{\mu\nu}$ at equilibrium is given by $\left<\bm{R}^{2}_{\mu\nu}\right>_{\text{eq}}\equiv\left<\bm{R}^{2}_{\mu\nu}(0)\right>=3|\nu-\mu|\left(k_BT/H\right)$. Using the length and time-scales given by $l_{H}=\sqrt{k_BT/H}$ and $\lambda_{H}=\zeta/4H$, respectively, and recognizing from Eq.~(\ref{eq:taup_def}) that
\begin{equation}\label{eq:dimless_conversion}
\dfrac{t}{\tau_p} \equiv\dfrac{H_pt}{\zeta_p}=\left(\dfrac{Ha_pt}{\zeta\left(1+\theta a_p\right)}\right)=\left(\dfrac{Ha_pt}{4H\lambda_{H}\left(1+\theta a_p\right)}\right)
=\left(\dfrac{1}{4}\right)\left(\dfrac{a_p}{1+\theta a_p}\right)\left(\dfrac{t}{\lambda_{H}}\right)=\left(\dfrac{a_p}{1+\theta a_p}\right)\dfrac{t^{*}}{4},
\end{equation}
the dimensionless form of Eq.~(\ref{eq:exp_nm_dim}) may be written as shown in Eq.~(\ref{eq:mu_nu_correl_in_paper}) of the main text.


\subsection{\label{sec:d_cont_rif_equiv} Equivalence between discrete and continuum versions of Rouse model with internal friction}

The autocorrelation of the end-to-end vector at equilibrium for the discrete model is given by Eq.~(\ref{eq:disc_exp_nm_dimless_main}). Recognizing that the time constant corresponding to each mode is given by Eq.~(\ref{eq:taup_def}), the expression for the autocorrelation may be written in terms of $t/\tau_1$ to be 
\begin{equation}\label{eq:exp_nm_scale_tau1}
\dfrac{\left<\bm{R}^{*}_{\text{E}}(0)\cdot\bm{R}^{*}_{\text{E}}(t/\tau_1)\right>}{\left<\bm{R}^{*2}_{\text{E}}(0)\right>}=\left[\dfrac{8}{N_{\mathrm{b}}\left(N_{\text{b}}-1\right)}\right]\,\sum_{p:\text{odd}}^{N_{\text{b}}-1}\cos^2\left(\dfrac{p\pi}{2N_{\mathrm{b}}}\right)\left(\dfrac{1}{a_p}\right)\exp\left[-\left(\dfrac{1+\theta a_1}{1+\theta a_p}\right)\left(\dfrac{a_p}{a_1}\right)\left(\dfrac{t}{\tau_1}\right)\right]
\end{equation}
\begin{figure}[t]
\begin{center}
\begin{tabular}{c}
\includegraphics[width=14cm,height=!]{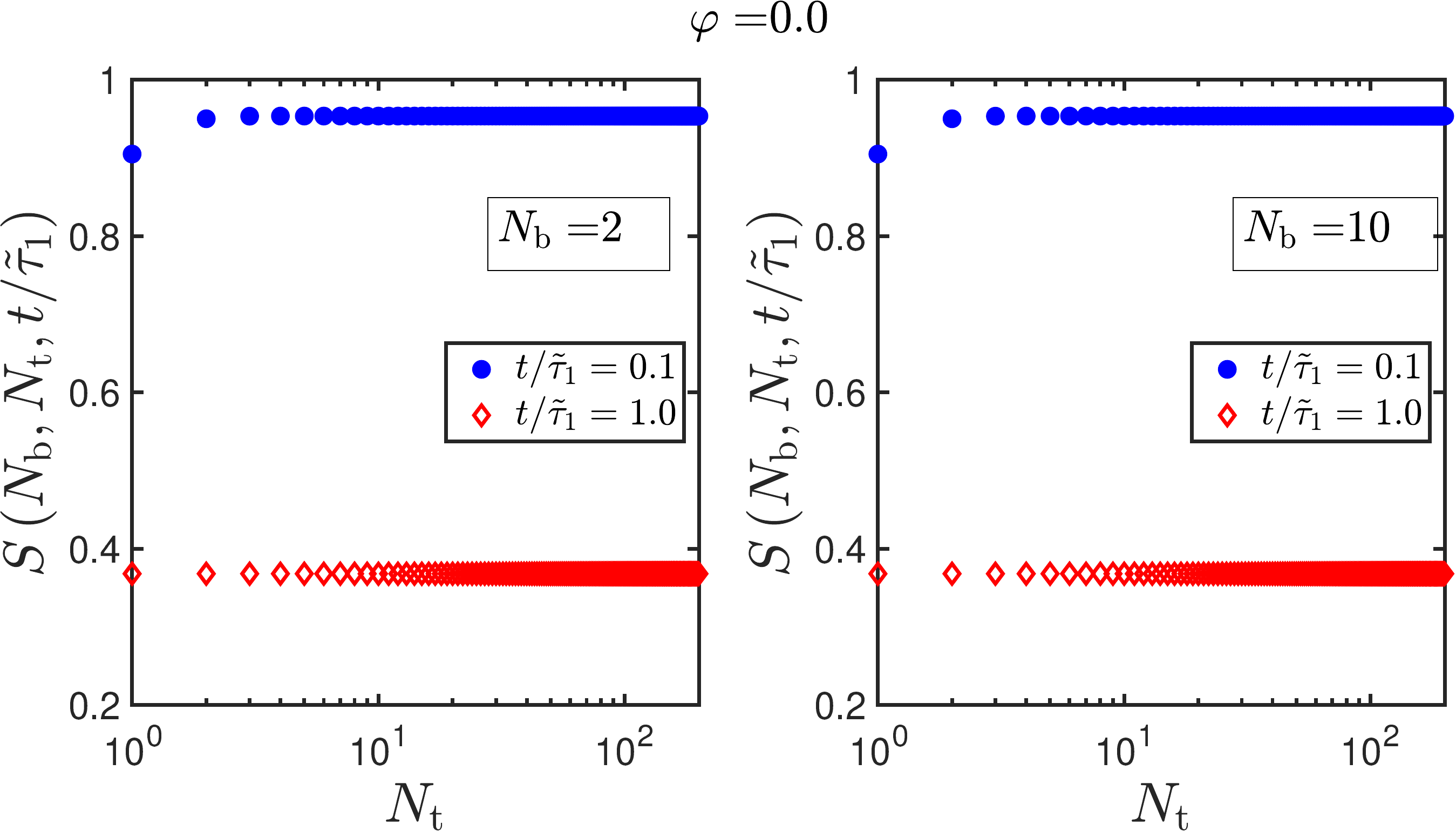}\\
(a)\\[10pt]
\includegraphics[width=14cm,height=!]{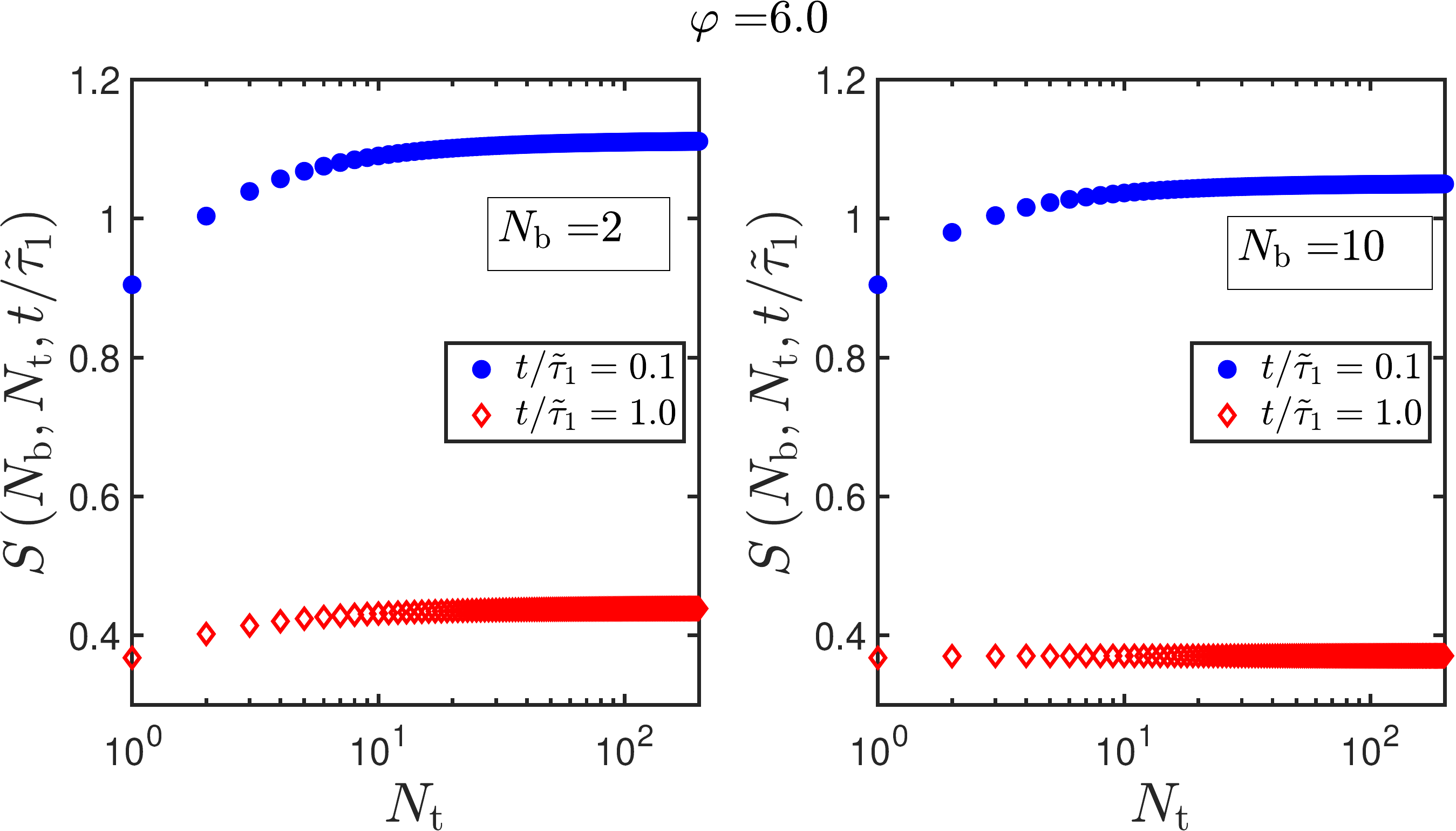}\\
(b)  \\
\end{tabular}
\end{center}
\vskip-10pt
\caption{\small Plot of number of terms required for convergence of summation indicated by Eq.~(\ref{eq:trunc_sum}) for two different chain lengths, for models with [(a)] and without [(b)] internal friction, at different values of the scaled time.}
\label{fig:sum_satur}
\end{figure}
\begin{figure*}[t]
\begin{center}
\begin{tabular}{c}
\includegraphics[width=16cm,height=!]{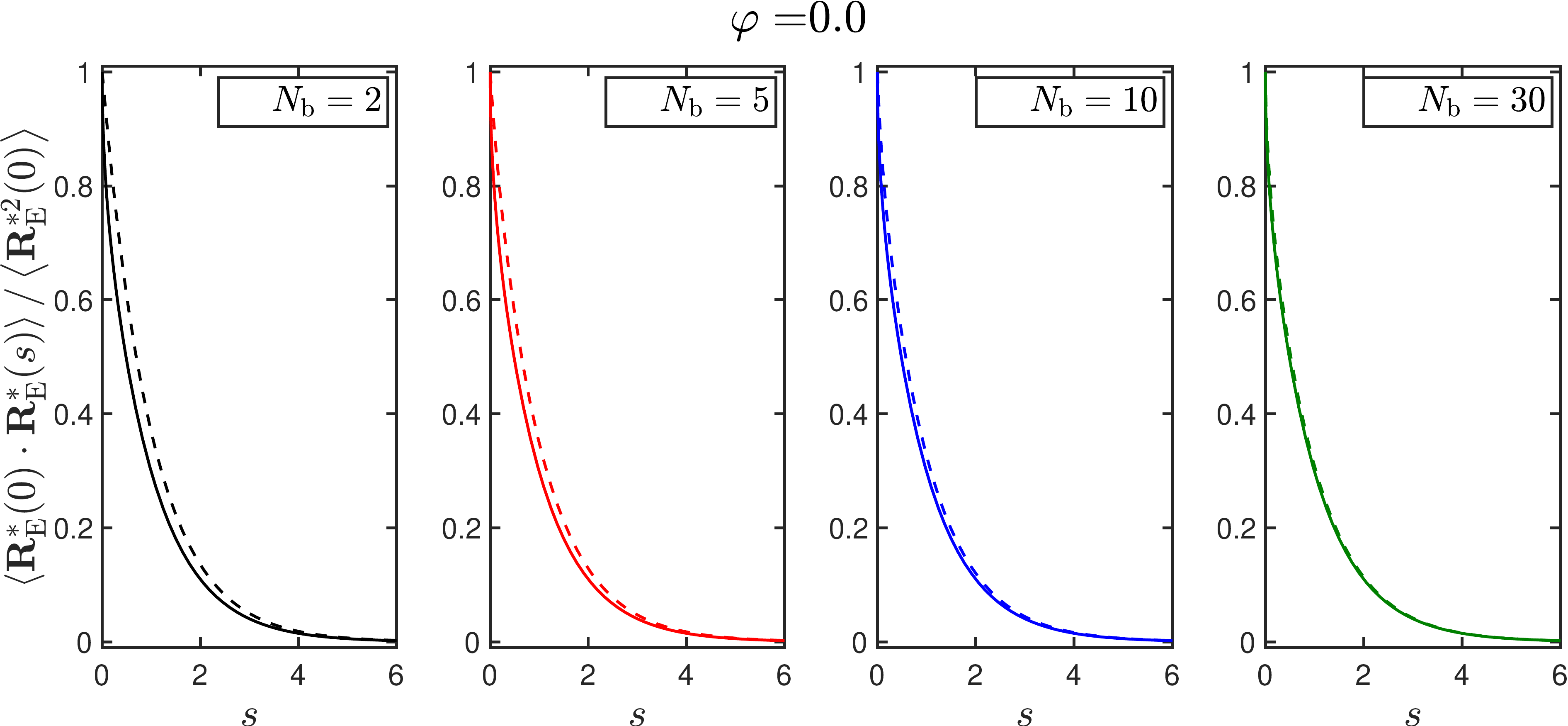}\\
(a)\\[10pt]
\includegraphics[width=16cm,height=!]{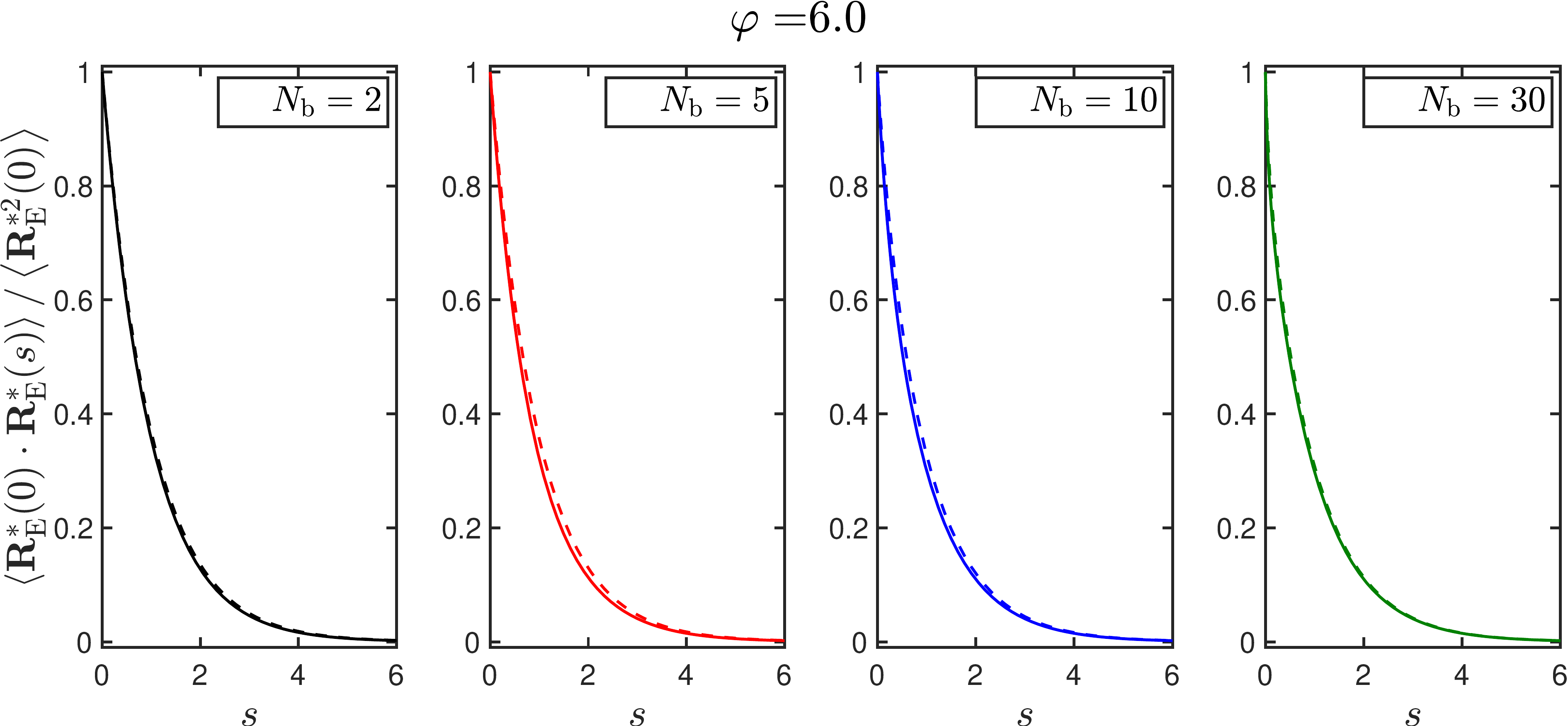}\\
(b)  \\
\end{tabular}
\end{center}
\vskip-10pt
\caption{Comparison of the normalized autocorrelation expressions for the discrete [dashed line, Eq.~(\ref{eq:exp_nm_scale_tau1})] and continuum [solid line, Eq.~(\ref{eq:cont_nm_scale_tau1}) ] RIF models, for various chain lengths, for models without [(a)] and with [(b)] internal friction. Note that $s\equiv t/\tau_1$ for the discrete model and $s\equiv t/\widetilde{\tau}_1$ for the continuum model.}
\label{fig:disc_cont_compare}
\end{figure*}
The normalized autocorrelation for the continuum RIF model is as follows~\cite{Khatri20071825,doi-edwards}
\begin{equation}\label{eq:cont_exp_nm_dim}
\dfrac{\left<\bm{R}^{*}_{\text{E}}(0)\cdot\bm{R}^{*}_{\text{E}}(t)\right>}{\left<\bm{R}^{*2}_{\text{E}}(0)\right>}=\left(\dfrac{8}{\pi^2}\right)\sum_{p:\text{odd}}^{\infty}\left(\dfrac{1}{p^2}\right)\exp\left[-\dfrac{t}{\widetilde{\tau}_p}\right],
\end{equation}
where 
\begin{equation}\label{eq:taup_tilde_def}
\widetilde{\tau}_p=\left(N^2_{\text{b}}\zeta/p^2\pi^2H\right)+ \left(K/3H\right), 
\end{equation}
and the dimensionless mean-squared end-to-end distance at equilibrium is given by $\left<\bm{R}^{*2}_{\text{E}}(0)\right>\equiv\left<\bm{R}^{*2}_{\text{E}}\right>_{\text{eq}}=3N_{\text{b}}$. 
The above expression may be rewritten in terms of $t/\widetilde{\tau}_1$ as
\begin{equation}\label{eq:cont_nm_scale_tau1}
\dfrac{\left<\bm{R}^{*}_{\text{E}}(0)\cdot\bm{R}^{*}_{\text{E}}(t/\widetilde{\tau}_1)\right>}{\left<\bm{R}^{*2}_{\text{E}}(0)\right>}=\left(\dfrac{8}{\pi^2}\right)\sum_{p:\text{odd}}^{\infty}\left(\dfrac{1}{p^2}\right)\exp\left\{-p^2\left[\dfrac{(N_{\mathrm{b}}/\pi)^2+\theta}{(N_{\mathrm{b}}/\pi)^2+p^2\theta}\right]\left(\dfrac{t}{\widetilde{\tau}_1}\right)\right\}
\end{equation}
From Eqs.~(\ref{eq:taup_def}) and~(\ref{eq:taup_tilde_def}), and recognizing that $\sin(x)\approx x$ as $x\to 0$, it is observed that $\tau_{p}\to\widetilde{\tau}_{p}$ as $N_{\text{b}}\to\infty$.
The infinite summation in Eq.~(\ref{eq:cont_nm_scale_tau1}) runs over all positive odd integers. We define a related quantity $S\left(N_{\text{b}},N_{\text{t}},t/\widetilde{\tau}_1\right)$, as
\begin{equation}\label{eq:trunc_sum}
S\left(N_{\text{b}},N_{\text{t}},t/\widetilde{\tau}_1\right)=\sum_{p=1,3,5,..}^{2N_{\text{t}}-1}\left(\dfrac{1}{p^2}\right)\exp\left\{-p^2\left[\dfrac{(N_{\mathrm{b}}/\pi)^2+\theta}{(N_{\mathrm{b}}/\pi)^2+p^2\theta}\right]\left(\dfrac{t}{\widetilde{\tau}_1}\right)\right\}
\end{equation}

\begin{figure*}[t]
\begin{center}
\includegraphics[width=17cm,height=!]{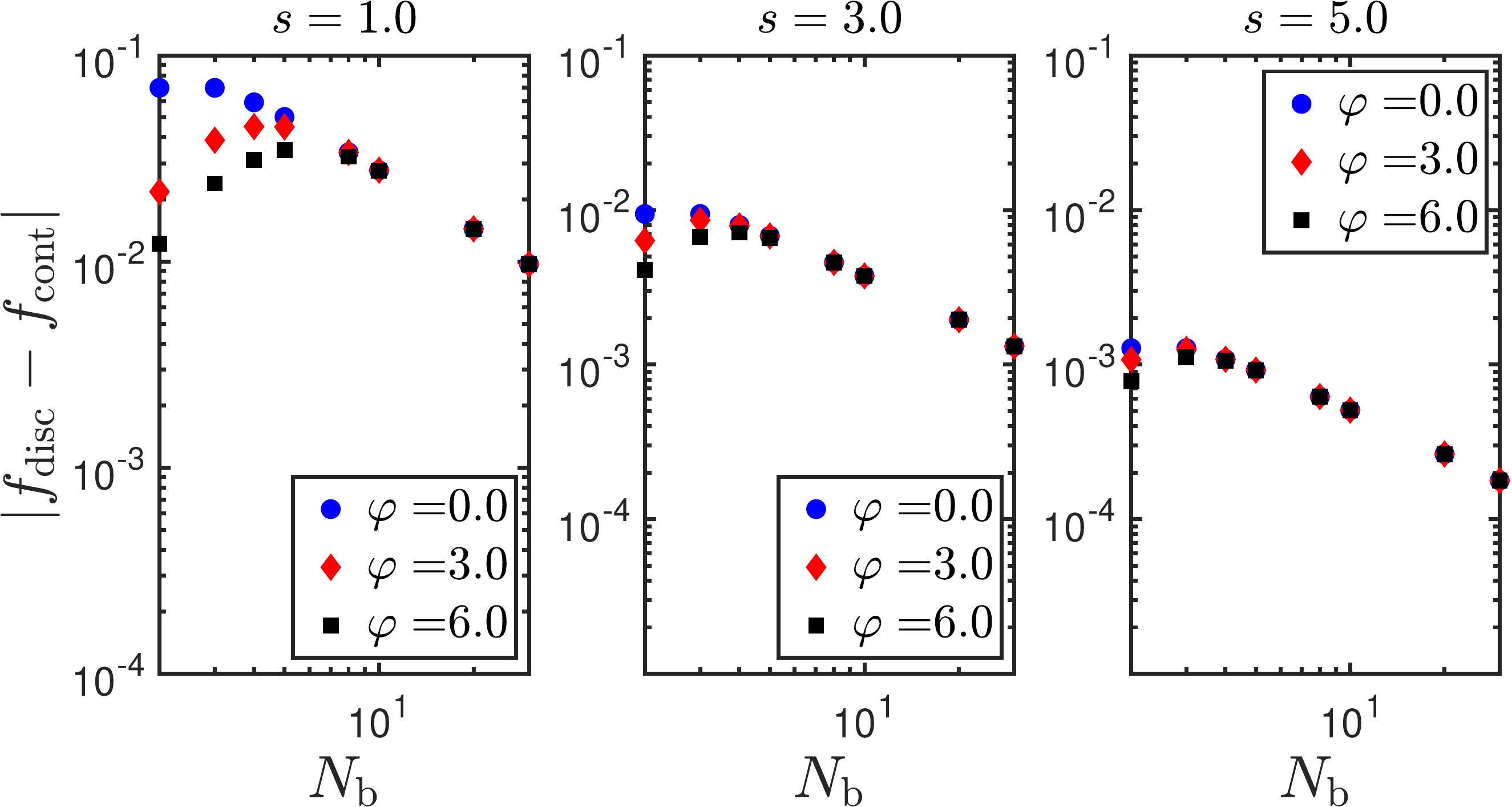}\\
\end{center}
\vskip-10pt
\caption{The absolute value of the difference between the discrete and continuum models at various instances of scaled time, plotted as a function of the chain length, for varying values of the internal friction parameter. Here, $f_{\text{disc}}$ represents the normalized autocorrelation for the discrete model at the indicated value of $s$, and $f_{\text{cont}}$ represents the normalized autocorrelation for the continuum model at the same value of $s$.}
\label{fig:compare_disc_cont}
\end{figure*}

\end{widetext}

In Fig.~\ref{fig:sum_satur}, the variation of $S$ as function of the number of terms included in the summation, $N_{\text{t}}$, for two different chain lengths, at two values of the dimensionless time and the internal friction parameter are displayed. In Fig.~\ref{fig:sum_satur}~(a), the case without internal friction is presented, and it is clearly seen that the summation requires fewer than ten terms for convergence at both early and later values of time. The summation at later times converges more quickly than the convergence at earlier times, for both two-bead and ten-bead chains. In Fig.~\ref{fig:sum_satur}~(b), the case with internal friction is presented. It is seen that nearly two hundred terms are required for the convergence of the sum for the two-bead chain at early times, and the corresponding number for the ten-bead chain at the same value of scaled time is marginally lower. As seen in (a), the summation at later times require fewer terms for convergence as compared to early times. We henceforth use two hundred terms in the numerical calculation of the infinite sum indicated in Eq.~(\ref{eq:cont_nm_scale_tau1}).

In Fig.~\ref{fig:disc_cont_compare}, it is seen that the normalized autocorrelation for the discrete model approaches the continuum result as the chain length is increased, for cases with and without internal friction. The difference is larger at early times, and lesser at later times, for all the values of chain length examined in the figure. In this figure, the normalized autocorrelation for the discrete chain is plotted as a function of time scaled by the longest relaxation time of the discrete chain [$t/\tau_1$], whereas the autocorrelation for the continuum model is plotted as a function of $t/\widetilde{\tau}_1$. The variable $s$ is used to refer to the scaled time, and its exact definition is context-dependent. For the case without internal friction [Fig.~\ref{fig:disc_cont_compare}~(a)], it is clearly seen that the difference between the discrete and continuum result decreases with an increase in the chain length.  However, for the case with internal friction [Fig.~\ref{fig:disc_cont_compare}~(b)], the difference appears to be non-monotonic in the chain length.

Fig.~\ref{fig:compare_disc_cont} examines the variation of the difference between the discrete and continuum result as a function of chain length, at three different instances of the scaled time $s$, for different values of the internal friction parameter. Note that the difference between the two models is taken at the same value of $s$. The magnitude of the difference is seen to be larger at shorter times, and smaller at later times, as previously seen in Fig.~\ref{fig:disc_cont_compare}, for models with and without internal friction. However, the nature of the variation of the difference with the chain length is significantly impacted by the presence of internal friction. For cases without internal friction, the difference decreases monotonically with the chain length. With the inclusion of internal friction, however, the difference hits a peak before decreasing monotonically with the chain length. The height of the peak is seen to be diminished at later times. 

\subsection{\label{sec:derv_shear_flow} Discrete RIF model in simple shear flow}

In this section, we derive an expression for the time evolution of the mean-squared end-to-end distance of a discrete RIF chain in simple shear flow. Transforming Eq.~(\ref{eq:bead_govern_with_shear}) into normal-mode coordinates leads to Eq.~(\ref{eq:mode_govern_with_shear}), which may be written in terms of the Cartesian components of $\bm{X}_p\equiv[x_p,y_p,z_p]^{T}$, and $\bm{g}_p\equiv\left[{g}^{(x)}_p,{g}^{(y)}_p,{g}^{(z)}_p\right]^{T}$ as 
\begin{equation}\label{eq:gov_cartes}
\dfrac{d}{dt}
\begin{bmatrix}
x_p\\
y_p\\
z_p
\end{bmatrix}=
-\left(\dfrac{1}{\tau_p}\right)
\begin{bmatrix}
x_p\\
y_p\\
z_p
\end{bmatrix}+
\left(\dfrac{1}{1+\theta a_p}\right)
\begin{bmatrix}
\dot{\gamma}y_p\\
0\\
0
\end{bmatrix}+
\begin{bmatrix}
{g}^{(x)}_p\\
{g}^{(y)}_p\\
{g}^{(z)}_p
\end{bmatrix}
\end{equation}
Recognizing that Eq.~(\ref{eq:gov_cartes}) represents a system of three linear stochastic differential equations, we write
\begin{align}\label{eq:ode_xp}
\dfrac{dx_p}{dt}+\left(\dfrac{1}{\tau_p}\right)x_p&=\left(\dfrac{\dot{\gamma}}{1+\theta a_p}\right)y_p + {g}^{(x)}_p
\end{align}
\begin{align}\label{eq:ode_yp}
\dfrac{dy_p}{dt}+\left(\dfrac{1}{\tau_p}\right)y_p&= {g}^{(y)}_p
\end{align}
\begin{align}\label{eq:ode_zp}
\dfrac{dz_p}{dt}+\left(\dfrac{1}{\tau_p}\right)z_p&={g}^{(z)}_p
\end{align}

The equation for $x_p$ depends explicitly on $y_p$, but $y_p$ is not coupled to $x_p$. Furthermore, $z_p$ evolves independently of $x_p$ and $y_p$. The methodology to solve for $y_p(t)$ is identical to that of solving for $z_p(t)$, and consequently, only the steps for the solution of $y_p(t)$ are given. The solution for $x_p(t)$ is dealt with subsequently. In the solution of these three equations [Eqs.~(\ref{eq:ode_xp})-(\ref{eq:ode_zp})], we closely follow the framework described in detail in Ref.~\citenum{Howard2011}. 

We note that
\begin{equation}\label{eq:prob_def_norm_mode}
\begin{split}
\left<\bm{X}_p(t)\cdot\bm{X}_q(t)\right>&=\Bigl<\left[x_p(t)\bm{e}_x+y_p(t)\bm{e}_y+z_p(t)\bm{e}_z\right] \Bigr. \\
& \cdot \Bigl. \left[x_q(t)\bm{e}_x+y_q(t)\bm{e}_y+z_q(t)\bm{e}_z\right]\Bigr>\\
=\left<x_p(t)x_q(t)\right> & +\left<y_p(t)y_q(t)\right>+\left<z_p(t)z_q(t)\right>
\end{split}
\end{equation}
and our task now involves the computation of the three ensemble-averaged quantities on the RHS of Eq.~(\ref{eq:prob_def_norm_mode}).
Starting from Eq.~(\ref{eq:ode_yp}), the formal solution for $y_p(t)$ is written as 
\begin{align}\label{eq:ysol_p}
y_p(t)&=y_p(0)e^{-t/\tau_p}+\int_{0}^{t}dt_1{g}^{(y)}_p(t_1)e^{-(t-t_1)/\tau_p}\nonumber\\[5pt]
&=y_p(0)e^{-t/\tau_p}+\Delta \mathcal{Y}_p(t)
\end{align}
The moments of $\Delta \mathcal{Y}_p(t)$ are
\begin{equation}\label{eq:zero_mean_gauss}
\begin{split}
\left<\Delta \mathcal{Y}_p(t)\right>&\equiv \left<\int_{0}^{t}dt_1{g}^{(y)}_p(t_1)e^{-(t-t_1)/\tau_p}\right>\\[5pt]
&=\int_{0}^{t}dt_1\left<{g}^{(y)}_p(t_1)\right>e^{-(t-t_1)/\tau_p}=0
\end{split}
\end{equation}
and
\begin{equation}\label{eq:del_y_corr}
\left<\Delta \mathcal{Y}_p(t)\Delta \mathcal{Y}_q(t)\right>=\dfrac{k_BT}{H_p}\delta_{pq}\left[1-e^{-2t/\tau_p}\right]
\end{equation}
The equal-time correlations of the $y-$ and $z-$ components are identical, and can be derived to be
\begin{equation}\label{eq:y_and_z_mode}
\left<y_p(t)y_q(t)\right>=\left<z_p(t)z_q(t)\right>=\dfrac{k_BT}{H_p}\delta_{pq}
\end{equation}
Starting from Eq.~(\ref{eq:ode_xp}), the formal solution for $x_p(t)$ is written as
\begin{equation}\label{eq:formal_xp}
\begin{split}
x_p(t)&=x_p(0)e^{-t/\tau_p}+\int_{0}^{t}dt_1\,{g}^{(x)}_p(t_1)e^{-(t-t_1)/\tau_p}\\[5pt]
&+\left(\dfrac{\dot{\gamma}}{1+\theta a_p}\right)\dashuline{\int_{0}^{t}dt_2\,y_p(t_2)e^{-(t-t_2)/\tau_p}}
\end{split}
\end{equation}
The underlined integral is evaluated as
\begin{widetext}
\begin{equation}\label{eq:mstep_begin}
\begin{split}
{\int_{0}^{t}dt_2\,y_p(t_2)e^{-(t-t_2)/\tau_p}}&=ty_p(0)e^{-t/\tau_p}+\int_{0}^{t}dt_2\int_{0}^{t_2}dt_3\,{g}^{(y)}_p(t_3)e^{-(t-t_3)/\tau_p}
\end{split}
\end{equation}
with the double-integral solved using the Cauchy formula for repeated integration~\cite{Oldhambook}, 
\begin{equation}\label{eq:cauchy_application}
\begin{split}
&\int_{0}^{t}dt_2\int_{0}^{t_2}dt_3\,{g}^{(y)}_p(t_3)e^{-(t-t_3)/\tau_p}=\dfrac{1}{\left(2-1\right)!}\int_{0}^{t}dt'\,\left(t-t'\right){g}^{(y)}_p(t')e^{-(t-t')/\tau_p},
\end{split}
\end{equation}
to obtain the following expression for $x_p(t)$,
\begin{equation}\label{eq:xsol_p}
\begin{split}
x_p(t)&=e^{-t/\tau_p}\left[x_p(0)+\left(\dfrac{\dot{\gamma}}{1+\theta a_p}\right)ty_p(0)\right]+\Delta \mathcal{X}_p^{(x)}(t) + \left(\dfrac{\dot{\gamma}}{1+\theta a_p}\right)\Delta \mathcal{X}_p^{(y)}(t)
\end{split}
\end{equation}
where
\begin{equation}\label{eq:x_comp_def}
\begin{split}
\Delta \mathcal{X}_p^{(x)}(t)&=\int_{0}^{t}dt_1{g}^{(x)}_p(t_1)e^{-(t-t_1)/\tau_p}\\
\Delta \mathcal{X}_p^{(y)}(t)&=\int_{0}^{t}dt'\left(t-t'\right){g}^{(y)}_p(t')e^{-(t-t')/\tau_p}
\end{split}
\end{equation}
The equal-time correlation of the $x-$ component is obtained as 
\begin{equation}\label{eq:correl_interme2}
\begin{split}
\left<x_p(t)x_q(t)\right>&=\left(\dfrac{k_BT}{H_p}\right)\delta_{pq}e^{-2t/\tau_p}+\uline{\left<\Delta \mathcal{X}_p^{(x)}(t)\Delta \mathcal{X}_q^{(x)}(t)\right>}+\left(\dfrac{\dot{\gamma}\tau_p}{1+\theta a_p}\right)^2\left(\dfrac{2t}{\tau_p}\right)^2\left(\dfrac{k_BT}{4H_p}\right)\delta_{pq}e^{-2t/\tau_p}\\[10pt]
&+\left(\dfrac{\dot{\gamma}}{1+\theta a_p}\right)\left(\dfrac{\dot{\gamma}}{1+\theta a_q}\right)\dashuline{\left<\Delta \mathcal{X}_p^{(y)}(t)\Delta \mathcal{X}_q^{(y)}(t)\right>}
\end{split}
\end{equation}
The solid-underlined term is identical to $\left<\Delta \mathcal{Y}_p(t)\Delta \mathcal{Y}_q(t)\right>$, and may be obtained from Eq.~(\ref{eq:del_y_corr}). The dashed underlined term may be simplified as 
\begin{equation}\label{eq:xcorr_intermed1}
\begin{split}
\left<\Delta \mathcal{X}_p^{(y)}(t)\Delta \mathcal{X}_q^{(y)}(t)\right>&=\left<\int_{0}^{t}dt'\left(t-t'\right){g}^{(y)}_p(t')e^{-(t-t')/\tau_p}\int_{0}^{t}dt''\left(t-t''\right){g}^{(y)}_p(t'')e^{-(t-t'')/\tau_q}\right>\\[5pt]
&=\dfrac{2k_BT}{\zeta_p}\delta_{pq}\int_{0}^{t}dt'\left(t-t'\right)^2e^{-2(t-t')/\tau_p}\\[5pt]
&=-\left(\dfrac{\tau_p}{2}\right)t^2e^{-2t/\tau_p}+\int_{0}^{t}\tau_p(t-t')e^{-2(t-t')/\tau_p}dt'\\[5pt]
&=-\left(\dfrac{\tau_p}{2}\right)t^2e^{-2t/\tau_p}+\left(\dfrac{\tau^2_p}{2}\right)\int_{0}^{t}(t-t')\dfrac{d}{dt'}\left[e^{-2(t-t')/\tau_p}\right]dt'\\[5pt]
&=\left(\dfrac{k_BT}{4H_p}\right)\tau^2_p\delta_{pq}\Biggl\{2-\left[2+2\left(\dfrac{2t}{\tau_p}\right)+\left(\dfrac{2t}{\tau_p}\right)^2\right]e^{-2t/\tau_p}\Biggr\}
\end{split}
\end{equation}
to give
\begin{equation}\label{eq:x_mode}
\begin{split}
\left<x_p(t)x_q(t)\right>&=\left(\dfrac{k_BT}{H_p}\right)\delta_{pq}+8\left(\dfrac{k_BT}{H_pa^2_p}\right)\delta_{pq}\left(\lambda_{H}\dot{\gamma}\right)^2\Biggl\{1-\left[1+\left(\dfrac{2t}{\tau_p}\right)\right]e^{-2t/\tau_p}\Biggr\}
\end{split}
\end{equation}
\noindent Plugging Eqs.~(\ref{eq:y_and_z_mode}) and (\ref{eq:x_mode}) into Eq.~(\ref{eq:prob_def_norm_mode}), the equal-time correlation for the modes is obtained as
\begin{equation}\label{eq:mode_result_pq}
\begin{split}
\left<\bm{X}_p(t)\cdot\bm{X}_q(t)\right>&=\left(\dfrac{3k_BT}{H}\right)\left(\dfrac{1}{a_p}\right)\delta_{pq}\Biggl\{1+\dfrac{8\left(\lambda_{\text{H}}\dot{\gamma}\right)^2}{3a_p^2}\left[1-\left(e^{-2t/\tau_p}\left[1+\left(\dfrac{2t}{\tau_p}\right)\right]\right)\right]\Biggr\}
\end{split}
\end{equation}
Recognizing that the end-to-end vector is given by
\begin{equation}
\begin{split}
\bm{R}_{\text{E}}(t)=-2\sqrt{\dfrac{2}{N_{\mathrm{b}}}}\,\,\sum_{q:\text{odd}}^{N_{\text{b}}-1}\cos\left(\dfrac{q\pi}{2N_{\mathrm{b}}}\right)\bm{X}_q(t),
\end{split}
\end{equation}
the expression for $\left<\bm{R}_{\text{E}}^{2}(t)\right>$ may be written as
\begin{equation}\label{eq:r_e_sq_exp}
\begin{split}
\left<\bm{R}_{\text{E}}^{2}(t)\right>&=\left(\dfrac{8}{N_{\text{b}}}\right)\sum_{q:\text{odd}}^{N_{\text{b}}-1}\sum_{p:\text{odd}}^{N_{\text{b}}-1}\cos\left(\dfrac{p\pi}{2N_{\mathrm{b}}}\right)\cos\left(\dfrac{q\pi}{2N_{\mathrm{b}}}\right)\left<\bm{X}_p(t)\cdot\bm{X}_q(t)\right>
\end{split}
\end{equation}
Using Eqs.~(\ref{eq:mode_result_pq}) and (\ref{eq:r_e_sq_exp}), we get
\begin{equation}\label{eq:r_e_sq_res}
\begin{split}
\left<\bm{R}_{\text{E}}^2(t)\right>&=\left(\dfrac{8}{N_{\mathrm{b}}}\right)\left(\dfrac{3k_BT}{H}\right)\,\sum_{p:\text{odd}}^{N_{\text{b}}-1}\left(\dfrac{1}{a_p}\right)\cos^2\left(\dfrac{p\pi}{2N_{\mathrm{b}}}\right)\Biggl\{1+\dfrac{8\left(\lambda_{\text{H}}\dot{\gamma}\right)^2}{3a_p^2}\left[1-\left(e^{-2t/\tau_p}\left[1+\left(\dfrac{2t}{\tau_p}\right)\right]\right)\right]\Biggr\},
\end{split}
\end{equation}
which may be written in the dimensionless form as
\begin{equation}\label{eq:trans_re_t}
\begin{split}
\left<\bm{R}_{\text{E}}^{*2}(t^*)\right>&=\left(\dfrac{24}{N_{\mathrm{b}}}\right)\,\sum_{p:\text{odd}}^{N_{\text{b}}-1}\left(\dfrac{1}{a_p}\right)\cos^2\left(\dfrac{p\pi}{2N_{\mathrm{b}}}\right)\Biggl\{1+\dfrac{8\left(\lambda_{\text{H}}\dot{\gamma}\right)^2}{3a_p^2}\Biggl[1-\Biggl(\exp\left[{-\left(\dfrac{a_p}{1+\theta a_p}\right)\dfrac{t^{*}}{2}}\right]\left[1+\left(\dfrac{a_p}{1+\theta a_p}\right)\dfrac{t^{*}}{2}\right]\Biggl)\Biggr]\Biggr\}
\end{split}
\end{equation}
The steady-state result is obtained by taking the limit $t^{*}\to\infty$ in Eq.~(\ref{eq:trans_re_t}), to give
\begin{equation}\label{eq:steady_re_t}
\begin{split}
\left<\bm{R}^{*2}_{\text{E}}\right>\equiv\left<\bm{R}_{\text{E}}^{*2}(t^*\to\infty)\right>&=\left(\dfrac{24}{N_{\mathrm{b}}}\right)\,\sum_{p:\text{odd}}^{N_{\text{b}}-1}\left(\dfrac{1}{a_p}\right)\cos^2\left(\dfrac{p\pi}{2N_{\mathrm{b}}}\right)\Biggl\{1+\dfrac{8\left(\lambda_{\text{H}}\dot{\gamma}\right)^2}{3a_p^2}\Biggr\}
\end{split}
\end{equation}
Scaling Eq.~(\ref{eq:trans_re_t}) by the mean-squared end-to-end vector leads to Eq.~(\ref{eq:nmlzd_trans_re_t}) in the main text.
\end{widetext}

Additionally, it is useful to compute the cross-correlation, $\left<x_p(t)y_q(t)\right>$, for subsequent use in the derivation of an analytical expression for the transient variation of the shear viscosity, as discussed later in Sec.~\ref{sec:derv_stress_tensor}. Note that this correlation is zero in the absence of flow, by virtue of the noise being white in normal mode space [cf. Eq.~\ref{eq:norm_mode_noise_rel}]. From Eq.~(\ref{eq:ysol_p}) and ~(\ref{eq:xsol_p}), we can write, after some algebra
\begin{equation}\label{eq:cross_correl_init}
\begin{split}
\left<x_p(t)y_q(t)\right> & =\left(\dfrac{\dot{\gamma}}{1+\theta\,a_{p}}\right)\delta_{pq}\left(\dfrac{k_BT}{H_p}\right)te^{-2t/\tau_{p}} \\[5pt] 
& +\left(\dfrac{\dot{\gamma}}{1+\theta\,a_{p}}\right)\dashuline{\left<\Delta \mathcal{X}_p^{(y)}(t)\Delta \mathcal{Y}_q(t)\right>}
\end{split}
\end{equation}
The underlined term is evaluated to be 
\begin{equation}\label{eq:del_cross_correl}
\begin{split}
 \left< \Delta \mathcal{X}_p^{(y)}(t)\Delta \mathcal{Y}_q(t)\right> = & \\[5pt]
 \left(\dfrac{k_BT}{H_p}\right)\left(\dfrac{\tau_p}{2}\right)\delta_{pq}\Biggl[1 & -e^{-2t/\tau_p}\left(1+\left(\dfrac{2t}{\tau_p}\right)\right)\Biggr]
\end{split}
\end{equation}
Plugging Eq.~(\ref{eq:del_cross_correl}) into Eq.~(\ref{eq:cross_correl_init}), we obtain
\begin{equation}
\begin{split}
\left<x_p(t)y_q(t)\right>&=\delta_{pq}\left(\dfrac{\dot{\gamma}}{1+\theta\,a_{p}}\right)\left(\dfrac{\tau_p}{2}\right)\left[1-e^{-2t/\tau_p}\right]
\end{split}
\end{equation}

\section{\label{sec:app_b} Equivalence between discrete RIF and the preaveraged internal friction model}

\subsection{\label{sec:derv_preav_sde} Stochastic differential equations}

We consider a freely-draining chain of $N_{\text{b}}$ massless beads where the neighbours are connected by means of a spring in parallel with a dashpot whose damping coefficient is $K$. The locations of the beads are given by $\{\bm{r}_{1},\bm{r}_2,\cdots,\bm{r}_{N_{\text{b}}}\}$, and the connector vector between adjacent beads is denoted by $\bm{Q}_j=\bm{r}_{j+1}-\bm{r}_j$, where $j\in[1,N]$. We assume an overdamped system that has equilibrated in momentum space, and the instantaneous normalized configuration distribution function for the chain is given by $\Psi\equiv\Psi\left(\bm{r}_1,\bm{r}_2,...,\bm{r}_{{N_{\text{b}}}},t\right)=\left(1/\mathcal{Z}\right)\exp\left[-\phi/k_BT\right]$, where $\phi$ denotes the intramolecular potential energy stored in the springs joining the beads and $\mathcal{Z}=\int\exp\left[-\phi/k_BT\right]d\bm{Q}_1d\bm{Q}_2\dots\,d\bm{Q}_{N}$. It is noted that $\Psi$ remains unmodified by the inclusion of internal friction. Using the principles of polymer kinetic theory~\cite{Bird1987b,ravibook,kailasham2021rouse}, the equation of motion for the momentum-averaged velocity of the $j^{\text{th}}$ connector vector may be derived to be
 \begin{equation}\label{eq:qdot}
 \begin{split}
 \llbracket\dot{\bm{Q}}_{j}\rrbracket&=\boldsymbol{\kappa}\cdot\bm{Q}_j-\dfrac{1}{\zeta}\sum_{k}A_{jk}\Biggl(k_BT\dfrac{\partial \ln \Psi}{\partial \bm{Q}_k}+\dfrac{\partial \phi}{\partial \bm{Q}_k}\\[5pt]
 &\qquad\qquad\qquad\qquad\quad+K\dashuline{\dfrac{\bm{Q}_k\bm{Q}_k}{\bm{Q}^2_k}}\cdot\llbracket\dot{\bm{Q}}_{k}\rrbracket\Biggr)
 \end{split}
 \end{equation}
 where
 \begin{align}\label{eq:rouse_matrix_def}
A_{jk}= \left\{
\begin{array}{ll}
       2; &  j=k \\[15pt]
      -1; & |j-k|=1 \\[15pt]
       0 ; & \text{otherwise}
\end{array} 
\right. 
\end{align}
Clearly, the equation for the $j^{\text{th}}$ connector-vector velocity is coupled to that of its nearest neighbours, which precludes not only the naive substitution of Eq.~(\ref{eq:qdot}) into an equation of continuity in $\Psi$, but also the derivation of the Fokker-Planck equation and the governing set of stochastic differential equations for the system, for all but the simplest case of a dumbbell ($N=1$). Manke and Williams~\cite{Manke1988} proposed a three-step iterative substitution methodology (details given below) for the decoupling of the connector vector velocities, and derived semi-analytical approximate expressions for the linear-viscoelastic properties~\cite{Manke1988,Dasbach1992} of chains with internal friction. In our recent work~\cite{kailasham2021rouse}, we have expanded the scope of the Manke and Williams decoupling methodology to obtain the governing set of exact stochastic differential equations for a Rouse chain with fluctuating internal friction that are valid both at equilibrium and in the presence of flow. The approximate solutions derived by Manke and Williams compare excellently against simulation results obtained by numerically integrating the exact stochastic differential equations using Brownian dynamics (BD) simulations for sufficiently long chains. We have also presented, for the first time, data on the steady-shear viscometric functions for Rouse chains with fluctuating internal friction, for the general case of $N>1$. The same methodology is applied here to solve for the governing stochastic differential equations of a Rouse chain with preaveraged internal friction.

As detailed in the discussion surrounding Eq.~(\ref{eq:iv_force_form_bead}) in the main text, the preaveraging approximation entails a replacement of the underlined term in Eq.~(\ref{eq:qdot}) by its average evaluated with respect to the equilibrium distribution function of a Rouse chain, which is $(\boldsymbol{\delta}/3)$. Therefore, for a chain with preaveraged internal friction, Eq.~(\ref{eq:qdot}) reduces to 
 \begin{equation}\label{eq:qdot_preav}
 \begin{split}
 \llbracket\dot{\bm{Q}}_{j}\rrbracket&=\boldsymbol{\kappa}\cdot\bm{Q}_j \\
 &- \dfrac{1}{\zeta}\sum_{k}A_{jk}\Biggl(k_BT\dfrac{\partial \ln \Psi}{\partial \bm{Q}_k}+\dfrac{\partial \phi}{\partial \bm{Q}_k}+\dfrac{K}{3}\llbracket\dot{\bm{Q}}_{k}\rrbracket\Biggr)
 \end{split}
 \end{equation}
 which may be simplified to give
 \begin{widetext}
\begin{equation}\label{eq:ck_gen_eq_final}
\begin{split}
 \llbracket\dot{\bm{Q}}_{j}\rrbracket&=\left(\dfrac{1}{1+2\theta}\right)\left(\boldsymbol{\kappa}\cdot\bm{Q}_k\right)-\left(\dfrac{k_BT}{\zeta}\right)\left(\dfrac{1}{1+2\theta}\right)\left[-\dfrac{\partial \ln \Psi}{\partial \bm{Q}_{k-1}}+2\dfrac{\partial \ln \Psi}{\partial \bm{Q}_k}-\dfrac{\partial \ln \Psi}{\partial \bm{Q}_{k+1}}\right]\\[5pt]
&-\left(\dfrac{1}{\zeta}\right)\left(\dfrac{1}{1+2\theta}\right)\left[-\dfrac{\partial \phi}{\partial \bm{Q}_{k-1}}+2\dfrac{\partial \phi}{\partial \bm{Q}_k}-\dfrac{\partial \phi}{\partial \bm{Q}_{k+1}}\right]+\left(\dfrac{\theta}{1+2\theta}\right) \llbracket\dot{\bm{Q}}_{j-1}\rrbracket+\left(\dfrac{\theta}{1+2\theta}\right) \llbracket\dot{\bm{Q}}_{j+1}\rrbracket
\end{split}
 \end{equation}
 An ensuing simplicity of the preaveraging approximation is that Eq.~(\ref{eq:ck_gen_eq_final}) may directly be subjected to the iterative-substitution-based decoupling methodology, unlike the case with fluctuations where the expression for $\bm{Q}_k\cdot\llbracket\dot{\bm{Q}}_{k}\rrbracket/{Q^2_{k}}$ must first be decoupled before obtaining the desired expression for $ \llbracket\dot{\bm{Q}}_{k}\rrbracket$.
  
Firstly, in the forward substitution step, the expression for $\llbracket\dot{\bm{Q}}_{j}\rrbracket$ is substituted into that for  $\llbracket\dot{\bm{Q}}_{j+1}\rrbracket$, iteratively, starting from $j=1$ until $j=(N-1)$. This results in the following general expression,
 \begin{equation}\label{eq:fwd_sub}
 \begin{split}
&\left(1-M_{k}\right) \llbracket\dot{\bm{Q}}_{k}\rrbracket=\left(\dfrac{\theta}{1+2\theta}\right) \llbracket\dot{\bm{Q}}_{k+1}\rrbracket+\left(\dfrac{1}{1+2\theta}\right)\sum_{l=1}^{k}{\Gamma}^{(k)}_{l}\left(\boldsymbol{\kappa}\cdot\bm{Q}_l\right)+\left(1-\delta_{kN}\right)\left(\dfrac{k_BT}{\zeta}\right)\left(\dfrac{1}{1+2\theta}\right)\left(\dfrac{\partial \ln \Psi}{\partial \bm{Q}_{k+1}}\right)\\[10pt]
&+\left(1-\delta_{kN}\right)\left(\dfrac{1}{\zeta}\right)\left(\dfrac{1}{1+2\theta}\right)\left(\dfrac{\partial \phi}{\partial \bm{Q}_{k+1}}\right)-\left(\dfrac{k_BT}{\zeta}\right)\left(\dfrac{1}{1+2\theta}\right)\sum_{l=1}^{k}{E}^{(k)}_{l}\left(\dfrac{\partial \ln \Psi}{\partial \bm{Q}_{l}}\right)-\left(\dfrac{1}{\zeta}\right)\left(\dfrac{1}{1+2\theta}\right)\sum_{l=1}^{k}{E}^{(k)}_{l}\left(\dfrac{\partial \phi}{\partial \bm{Q}_{l}}\right)
\end{split}
\end{equation}
where the explicit dependence of $\llbracket\dot{\bm{Q}}_{k}\rrbracket$ on  $\llbracket\dot{\bm{Q}}_{k-1}\rrbracket$ has been removed and the following definitions apply
\begin{equation}\label{eq:m_def}
\begin{split}
M_{k}=\left(\dfrac{\theta}{1+2\theta}\right)^2\left(\dfrac{1}{1-M_{k-1}}\right);\,\text {with} \quad M_{1}=0
\end{split}
\end{equation}
\begin{equation}\label{eq:gamma_def}
{\Gamma}^{(k)}_l=\left(\dfrac{\theta}{1+2\theta}\right)^{k-l}\,\,\,\prod_{i=l}^{k-1}\left(\dfrac{1}{1-M_i}\right)
\end{equation}
\begin{equation}\label{eq:e_def}
{E}^{(k)}_l=2{\Gamma}^{(k)}_l-{\Gamma}^{(k)}_{l-1}-{\Gamma}^{(k)}_{l+1}
\end{equation}
Next, in the backward substitution step, the expression for $\llbracket\dot{\bm{Q}}_{j}\rrbracket$ is substituted into that for  $\llbracket\dot{\bm{Q}}_{j-1}\rrbracket$, iteratively, starting from $j=N$ until $j=2$. This results in the following general expression,
\begin{equation}\label{eq:bkwd_sub_intermed}
 \begin{split}
&\llbracket\dot{\bm{Q}}_{k}\rrbracket=\left(\dfrac{\theta}{1+2\theta}\right)\left(\dfrac{1}{1-P_{k}}\right)\llbracket\dot{\bm{Q}}_{k-1}\rrbracket+\left(\dfrac{1}{1+2\theta}\right)\sum_{l=k}^{N}\widetilde{\rho}^{(k)}_{l}\left(\boldsymbol{\kappa}\cdot\bm{Q}_l\right)+\left(\dfrac{k_BT}{\zeta}\right)\left(\dfrac{1}{1+2\theta}\right)\left(\dfrac{1}{1-P_{k}}\right)\left(\dfrac{\partial \ln \Psi}{\partial \bm{Q}_{k-1}}\right)\\[10pt]
&+\left(\dfrac{1}{\zeta}\right)\left(\dfrac{1}{1+2\theta}\right)\left(\dfrac{1}{1-P_{k}}\right)\left(\dfrac{\partial \phi}{\partial \bm{Q}_{k-1}}\right)-\left(\dfrac{k_BT}{\zeta}\right)\left(\dfrac{1}{1+2\theta}\right)\sum_{l=k}^{N}\widetilde{G}^{(k)}_{l}\left(\dfrac{\partial \ln \Psi}{\partial \bm{Q}_{l}}\right)-\left(\dfrac{1}{\zeta}\right)\left(\dfrac{1}{1+2\theta}\right)\sum_{l=k}^{N}\widetilde{G}^{(k)}_{l}\left(\dfrac{\partial \phi}{\partial \bm{Q}_{l}}\right)
\end{split}
\end{equation}
\end{widetext}
where the explicit dependence of $\llbracket\dot{\bm{Q}}_{k}\rrbracket$ on  $\llbracket\dot{\bm{Q}}_{k+1}\rrbracket$ has been removed and the following definitions apply
\begin{equation}\label{eq:p_def}
\begin{split}
P_{k}=\left(\dfrac{\theta}{1+2\theta}\right)^2\left(\dfrac{1}{1-P_{k+1}}\right); \,\text {with} \quad P_{N}=0.
\end{split}
\end{equation}
\begin{equation}\label{eq:rho_tilde_def}
\widetilde{\rho}^{(k)}_{l}=\left(\dfrac{\theta}{1+2\theta}\right)^{l-k}\,\,\,\prod_{i=k}^{l}\left(\dfrac{1}{1-P_i}\right)
\end{equation}

The quantity $\widetilde{{G}}^{(k)}_l$ appearing in Eq.~(\ref{eq:bkwd_sub_intermed}) is constructed using a slightly elaborate procedure. It is useful to first consider the matrix $\bm{A}$ [defined as in Eq.~(\ref{eq:rouse_matrix_def})], of size $\Upsilon\times\Upsilon$, where $\Upsilon=\left(N-k\right)+1$, and the intermediate quantity, 
\begin{align}\label{eq:y_tilde_def}
\widetilde{{Y}}^{(k)}_{s}&=\left(\dfrac{\theta}{1+2\theta}\right)^{s-1}\left[\prod_{i=k}^{k+s-1}\left(\dfrac{1}{1-P_i}\right)\right]
\end{align}
which is then used to populate a matrix, $\widehat{\boldsymbol{\Theta}}^{(k)}$, of size $\Upsilon\times\Upsilon$ that has the following structure
\begin{equation}\label{eq:theta_hat_def}
\widehat{\boldsymbol{\Theta}}^{(k)} = 
\begin{pmatrix}
\widetilde{{Y}}^{(k)}_{1} & \widetilde{{Y}}^{(k)}_{1} & {0} & \cdots & {} & {}\\
\widetilde{{Y}}^{(k)}_{2}  & \widetilde{{Y}}^{(k)}_{2}  & \widetilde{{Y}}^{(k)}_{2}  &{0} & \cdots & {} \\
{0} & \widetilde{{Y}}^{(k)}_{3} & \widetilde{{Y}}^{(k)}_{3} & \widetilde{{Y}}^{(k)}_{3} &\cdots & {} \\
\vdots  & \vdots & \vdots & {}& {} & {} \\
{0} & {0}& \cdots & {} & \widetilde{{Y}}^{(k)}_{\Upsilon}& \widetilde{{Y}}^{(k)}_{\Upsilon}
\end{pmatrix}
\end{equation}

We next consider the matrix $\bm{Z}^{(k)}$ constructed from $\bm{A}$ [see Eq.~(\ref{eq:rouse_matrix_def})] and $\widehat{\boldsymbol{\Theta}}^{(k)}$, such that $\bm{Z}^{(k)}=\bm{A}\cdot\widehat{\boldsymbol{\Theta}}^{(k)}$. Now, $\widetilde{{G}}^{(k)}_{k+m}$ is defined as the $\left(m+1\right)^{\text{th}}$ diagonal element of $\bm{Z}^{(k)}$.  As the final step of the decoupling procedure, a change of variable, $k\to\left(k+1\right)$, is performed in Eq.~(\ref{eq:bkwd_sub_intermed}), and the resulting expression is substituted into the equation derived from the forward substitution step [Eq.~(\ref{eq:fwd_sub})]. The decoupled expression for $\llbracket\dot{\bm{Q}}_{k}\rrbracket$ is finally obtained as 
\begin{widetext}
\begin{equation}\label{eq:decoup_init}
\begin{split}
\llbracket\dot{\bm{Q}}_{k}\rrbracket&=\left(\dfrac{1}{1+2\theta}\right)\sum_{l=1}^{N}{{\Lambda}}_{kl}\left(\boldsymbol{\kappa}\cdot\bm{Q}_l\right)-\left(\dfrac{1}{\zeta}\right)\left(\dfrac{1}{1+2\theta}\right)\sum_{l=1}^{N}{{J}}_{kl}\left(\dfrac{\partial \phi}{\partial \bm{Q}_{l}}\right)-\left(\dfrac{k_BT}{\zeta}\right)\left(\dfrac{1}{1+2\theta}\right)\sum_{l=1}^{N}{{J}}_{kl}\left(\dfrac{\partial \ln \Psi}{\partial \bm{Q}_{l}}\right)
\end{split}
\end{equation}
In defining Eqs.~(\ref{eq:gamma_def}), (\ref{eq:rho_tilde_def}), and (\ref{eq:y_tilde_def}), we have adopted the convention $0^{0}=1$, i.e., when $\theta=0$, ${\Gamma}^{(k)}_k=\widetilde{\rho}^{(k)}_{k}=\widetilde{{Y}}^{(k)}_{1}=1$. The matrix elements, ${{\Lambda}}_{kl}$, and ${{J}}_{kl}$, each of size $N\times N$, are defined as
\begin{equation}\label{eq:lambda_j_def}
\begin{split}
{{\Lambda}}_{kl}=\left(\dfrac{1}{1-M_k-P_k}\right)\widehat{{\Lambda}}_{kl};\quad {{J}}_{kl}&=\left(\dfrac{1}{1-M_k-P_k}\right)\widehat{{J}}_{kl}
\end{split}
\end{equation}
with
\begin{align}\label{eq:flow_coeffdef}
\widehat{{\Lambda}}_{kl}= \left\{
\begin{array}{ll}
       {\Gamma}^{(k)}_l; &  l<k\\[15pt]
       1; & l=k\\[15pt]
       \left(\dfrac{\theta}{1+2\theta}\right)\widetilde{\rho}^{(k+1)}_{l} ; & l>k
\end{array} 
\right. 
\end{align}
and
\begin{align}\label{eq:stiff_coeffdef}
\widehat{{J}}_{kl}= \left\{
\begin{array}{ll}
       {E}^{(k)}_l; &  l<k\\[15pt]
       {E}^{(k)}_l - \left(1-\delta_{kN}\right)\left(\dfrac{1}{1-P_{k+1}}\right)\left(\dfrac{\theta}{1+2\theta}\right); & l=k\\[15pt]
        \left(1-\delta_{kN}\right)\left[ \left(\dfrac{\theta}{1+2\theta}\right)\widetilde{G}^{(k+1)}_{l}-1\right] ; & l=k+1\\[15pt]
        \left(\dfrac{\theta}{1+2\theta}\right)\widetilde{G}^{(k+1)}_{l} ; & l>\left(k+1\right)
\end{array} 
\right. 
\end{align}
\end{widetext}

The procedure for the construction of $\widetilde{G}^{(k+1)}_{l}$ which appears in Eq.~(\ref{eq:stiff_coeffdef}) is fairly similar to that described in Eq.~(\ref{eq:theta_hat_def}) for the construction of $\widetilde{{G}}^{(k)}_l$, with the only caveat that the size of the block matrices, $\bm{A}$ and the $\bm{\widehat{\Theta}}^{(k+1)}$, remain $\Upsilon\times\Upsilon$, where $\Upsilon=\left(N-k\right)+1$.

Another point of difference between the preaveraged IV model and the fluctuating IV one is that for the former, the quantities $\{M_k,P_k,{\Gamma}^{(k)}_l,E^{(k)}_l,\widetilde{\rho}^{(k)}_{l},\widetilde{{Y}}^{(k)}_{s},\widetilde{{G}}^{(k)}_l,{{\Lambda}}_{kl}, J_{kl}\}$ are functions only of the internal friction parameter $\theta$, and not dependent on the chain configuration. In the fluctuating IV model, however, these quantities are functions of both the internal friction parameter, and the chain configuration.

As the next step, the expression for $ \llbracket\dot{\bm{Q}}_{k}\rrbracket$ will be substituted into the equation of continuity, recognizing that the homogeneous flow profile allows one to write the continuity equation solely in terms of the relative coordinates, $\bm{Q}_k$. This means that the distribution function $\Psi\left(\bm{r}_c,\bm{Q}_1,\bm{Q}_2,...\bm{Q}_N\right)$ can be replaced by $\psi\left(\bm{Q}_1,\bm{Q}_2,...\bm{Q}_N\right)$, and we have
\begin{equation}
\begin{split}
\dfrac{\partial \psi}{\partial t}&=-\sum_{k=1}^{N}\dfrac{\partial}{\partial \bm{Q}_k}\cdot\left\{\llbracket\dot{\bm{Q}}_{k}\rrbracket\psi\right\}\\[10pt]
&=-\sum_{k=1}^{N}\dfrac{\partial}{\partial \bm{Q}_k}\cdot\Biggl\{\Biggl[\left(\dfrac{1}{1+2\theta}\right)\sum_{l=1}^{N}{\boldsymbol{\Lambda}}_{kl}\cdot\left(\boldsymbol{\kappa}\cdot\bm{Q}_l\right)\\[10pt]
&-\left(\dfrac{1}{\zeta}\right)\left(\dfrac{1}{1+2\theta}\right)\sum_{l=1}^{N}{\bm{J}}_{kl}\cdot\left(\dfrac{\partial \phi}{\partial \bm{Q}_{l}}\right)\Biggr]\psi\Biggr\}\\[10pt]
&+\left(\dfrac{k_BT}{\zeta}\right)\left(\dfrac{1}{1+2\theta}\right)\sum_{k=1}^{N}\sum_{l=1}^{N}\dfrac{\partial}{\partial \bm{Q}_k}\cdot\left[\bm{J}_{kl}\cdot\dfrac{\partial \psi}{\partial \bm{Q}_{l}}\right]
\end{split}
\end{equation}
where $\bm{J}_{kl}=J_{kl}\boldsymbol{\delta}$, and $\bm{\Lambda}_{kl}=\Lambda_{kl}\boldsymbol{\delta}$. 
Since $\bm{J}_{kl}$ is composed entirely of constant coefficients that are independent of the stochastic variables $\bm{Q}_k$, it is divergence-free, and the noise term may be rewritten, giving the following Fokker-Planck equation
\begin{equation}\label{eq:fokker_planck}
\begin{split}
\dfrac{\partial \psi}{\partial t}&=-\sum_{k=1}^{N}\dfrac{\partial}{\partial \bm{Q}_k}\cdot\Biggl\{\Biggl[\left(\dfrac{1}{1+2\theta}\right)\sum_{l=1}^{N}{\boldsymbol{\Lambda}}_{kl}\cdot\left(\boldsymbol{\kappa}\cdot\bm{Q}_l\right)\\[10pt]
&-\left(\dfrac{1}{\zeta}\right)\left(\dfrac{1}{1+2\theta}\right)\sum_{l=1}^{N}{\bm{J}}_{kl}\cdot\left(\dfrac{\partial \phi}{\partial \bm{Q}_{l}}\right)\Biggr]\psi\Biggr\}\\[10pt]
&+\left(\dfrac{k_BT}{\zeta}\right)\left(\dfrac{1}{1+2\theta}\right)\sum_{k=1}^{N}\sum_{l=1}^{N}\dfrac{\partial}{\partial \bm{Q}_k}\dfrac{\partial }{\partial \bm{Q}_{l}}:\left[\bm{J}_{kl}\psi\right]
\end{split}
\end{equation}
where we have implicitly used the fact that $\bm{J}^{T}_{kl}={J}_{kl}\boldsymbol{\delta}^{T}=\bm{J}_{kl}$, and $J_{kl}=J_{lk}$.  Noting that $\bm{F}_l^{(\phi)}\equiv\left(\partial \phi/\partial \bm{Q}_{l}\right)=H\bm{Q}_l$ for Hookean springs, the stochastic differential equation may be written, using the It\^o interpretation~\cite{Ottinger1996} of Eq.~(\ref{eq:fokker_planck}), as
\begin{equation}\label{eq:sde_individual}
\begin{split}
d\bm{Q}_{k}&=\left(\dfrac{1}{1+2\theta}\right)\Biggl[\sum_{l=1}^{N}{{\Lambda}}_{kl}\left(\boldsymbol{\kappa}\cdot\bm{Q}_l\right)\\[10pt]
&-\left(\dfrac{H}{\zeta}\right)\sum_{l=1}^{N}{{J}}_{kl}\bm{Q}_l\Biggr]dt+\sqrt{\dfrac{2k_BT}{\zeta(1+2\theta)}}\sum_{l=1}^{N}{B}_{kl}\,d\bm{W}_{l}
\end{split}
\end{equation}
where
\begin{equation}\label{eq:sqrt_tens}
\sum_{j=1}^{N}{B}_{kj}{B}_{lj}={J}_{kl}
\end{equation}
Finally, the governing equation in its dimensionless form is given by
\begin{equation}\label{eq:sde_individual_dimless}
\begin{split}
d\bm{Q}^{*}_{k}&=\left(\dfrac{1}{1+2\theta}\right)\Biggl[\sum_{l=1}^{N}{{\Lambda}}_{kl}\left(\boldsymbol{\kappa}^{*}\cdot\bm{Q}^{*}_l\right)\\[10pt]
&-\left(\dfrac{1}{4}\right)\sum_{l=1}^{N}{{J}}_{kl}\bm{Q}^{*}_l\Biggr]dt^{*}+\sqrt{\dfrac{1}{2(1+2\theta)}}\sum_{l=1}^{N}{B}_{kl}\,d\bm{W}^{*}_{l},
\end{split}
\end{equation}
Eq.~(\ref{eq:sde_individual_dimless}) is integrated numerically using the simple Euler discretization~\cite{Ottinger1996} method, with a time-step width of $\Delta t^{*}=10^{-3}$. Averages are evaluated over an ensemble of $\mathcal{O}(10^5)$ trajectories. Since Eq.~(\ref{eq:sde_individual}) is a linear stochastic differential equation, it can also be solved semi-analytically, as described in the section below.

\vspace{20pt}

\subsection{\label{sec:equivalence} Semi-analytical solution to the preaveraged internal friction model}

The Langevin equation corresponding to Eq.~(\ref{eq:sde_individual}) is given by
\begin{equation}\label{eq:langevin_general}
\begin{split}
\dfrac{d\bm{Q}_{k}}{dt} & = \left(\dfrac{1}{1+2\theta}\right)\Biggl[\sum_{l=1}^{N}{{\Lambda}}_{kl}\left(\boldsymbol{\kappa}\cdot\bm{Q}_l\right)-\left(\dfrac{H}{\zeta}\right)\sum_{l=1}^{N}{{J}}_{kl}\bm{Q}_{l}\Biggr] \\[10pt] 
& + \sqrt{\dfrac{2k_BT}{\zeta(1+2\theta)}}\sum_{l=1}^{N}{B}_{kl}\,\bm{f}_{l}(t)
\end{split}
\end{equation}
where
\begin{equation}\label{eq:noise_prop}
\begin{split}
\left<\bm{f}_l(t)\right>&=0\\[5pt]
\left<\bm{f}_l(t)\bm{f}_m(t')\right>&=\delta_{ml}\boldsymbol{\delta}\delta(t-t')\\[5pt]
\left<\bm{f}_l(t)\cdot \bm{f}_m(t')\right>&=3\,\delta_{ml}\,\delta(t-t')
\end{split}
\end{equation}
Semi-analytical solutions to Eq.~(\ref{eq:langevin_general}) at equilibrium and in the presence of shear flow are presented below.

 \subsubsection{\label{sec:derv_preav_eqbm} Normal mode analysis at equilibrium}

To obtain the correlation of the end-to-end vector for a Rouse chain with preaveraged internal friction, we set $\boldsymbol{\kappa}=0$ in Eq.~(\ref{eq:langevin_general}) and obtain the governing Langevin equation as 
\begin{equation}\label{eq:langevin_noflow}
\dfrac{d\bm{Q}_{k}}{dt}  =  -\left[\dfrac{H}{\zeta(1+2\theta)}\right]\sum_{l=1}^{N}{{J}}_{kl}\bm{Q}_l 
 +\sqrt{\dfrac{2k_BT}{\zeta(1+2\theta)}}\sum_{l=1}^{N}{B}_{kl}\,\bm{f}_{l}(t)
\end{equation}
This equation is converted into normal mode coordinates by means of the transformation $\bm{Q}'_j=\sum_{m}\Pi_{mj}\bm{Q}_m$, where the orthogonalizing matrix $\boldsymbol{\Pi}$ satisfies the following properties
\begin{equation}\label{eq:eig_decomp}
\begin{split}
&\sum_{k=1}^{N}\Pi_{kj}\Pi_{kl}=\delta_{jl}=\sum_{n=1}^{N}\Pi_{jn}\Pi_{ln}\\[10pt]
&\sum_{l=1}^{N}\sum_{n=1}^{N}\Pi_{lj}J_{ln}\Pi_{nk}=\widetilde{a}_{j}\delta_{jk}
\end{split}
\end{equation}
and must be determined numerically. In Eq.~(\ref{eq:eig_decomp}), the $\widetilde{a}_j$ represent the eigenvalues of $\bm{J}$. 

\begin{widetext}
The governing equation for the normal coordinates is given as
\begin{equation}\label{eq:norm_mode_diff_eq}
\dfrac{d\bm{Q}'_{m}}{dt} =-\left[\dfrac{H\widetilde{a}_m}{\zeta(1+2\theta)}\right]\bm{Q}'_{m} 
  +\sqrt{\dfrac{2k_BT}{\zeta(1+2\theta)}}\sum_{k}\sum_{l}\Pi_{km}{B}_{kl}\,\bm{f}_{l}(t)
\end{equation}
whose formal solution may be written as
\begin{equation}\label{eq:q_m}
\begin{split}
&\bm{Q}'_{m}(t)=\bm{Q}'_{m}(0)e^{-t/\widehat{\tau}_m}+\sqrt{\dfrac{2k_BT}{\zeta(1+2\theta)}}\int_{0}^{t}dt_{1}\left[\sum_{k}\sum_{l}\Pi_{km}{B}_{kl}\,\bm{f}_{l}(t_1)\right]e^{-(t-t_1)/\widehat{\tau}_m}
\end{split}
\end{equation}
where $\widehat{\tau}_m={\zeta(1+2\theta)}/{H\widetilde{a}_m}$. The autocorrelation of the normal modes may be derived to be
\begin{equation}\label{eq:mode_soln}
\left<\bm{Q}'_{p}(0)\cdot \bm{Q}'_{m}(t)\right>=\dfrac{3k_BT}{H}\delta_{pm}e^{-t/\widehat{\tau}_p}
\end{equation}
Recognizing that
\begin{equation}\label{eq:re_intermed}
\left<\bm{R}_{\text{E}}(0)\cdot\bm{R}_{\text{E}}(t)\right>=\sum_{i,m,j,p}\Pi_{im}\Pi_{jp}\left<\bm{Q}'_p(0)\cdot\bm{Q}'_m(t)\right>
\end{equation}
and
\begin{equation}\label{eq:dimless_conversion_hat}
\begin{split}
\dfrac{t}{\widehat{\tau}_p}&\equiv \left[\dfrac{H\widetilde{a}_pt}{\zeta\left(1+2\theta\right)}\right]=\left[\dfrac{H\widetilde{a}_pt}{4H\lambda_{H}\left(1+2\theta\right)}\right]=\left(\dfrac{1}{4}\right)\left(\dfrac{\widetilde{a}_p}{1+2\theta}\right)\left(\dfrac{t}{\lambda_{H}}\right)=\left(\dfrac{\widetilde{a}_p}{1+2\theta}\right)\dfrac{t^{*}}{4}
\end{split}
\end{equation}
the normalized autocorrelation of the end-to-end vector in dimensionless time units may be written as
\begin{equation}\label{eq:preav_disc_exp_nm_dimless}
\begin{split}
\dfrac{\left<\bm{R}^{*}_{\text{E}}(0)\cdot\bm{R}^{*}_{\text{E}}(t^*)\right>}{\left<\bm{R}^{*2}_{\text{E}}(0)\right>}=\left[\dfrac{1}{N_{\text{b}}-1}\right]&\sum_{i,j,p}^{N_{\text{b}}-1}\Pi_{ip}\Pi_{jp}\exp\left[-\left(\dfrac{\widetilde{a}_p}{1+2\theta}\right)\dfrac{t^{*}}{4}\right]
\end{split}
\end{equation} 

 \subsubsection{\label{sec:derv_preav_shear} Normal mode analysis in shear flow}
 
The governing Langevin equation for a Rouse chain with preaveraged internal friction in shear flow is written as
 \begin{equation}\label{eq:langevin_shear_flow}
\begin{split}
\dfrac{d\bm{Q}_{k}}{dt}&=\left(\dfrac{1}{1+2\theta}\right)\left[\sum_{l=1}^{N}\Lambda_{kl}\left(\boldsymbol{\kappa}\cdot\bm{Q}_{l}\right)-\dfrac{H}{\zeta}{{J}}_{kl}\bm{Q}_l\right]+\sqrt{\dfrac{2k_BT}{\zeta(1+2\theta)}}\sum_{l=1}^{N}{B}_{kl}\,\bm{f}_{l}(t)
\end{split}
\end{equation}
Eq.~(\ref{eq:langevin_shear_flow}) is transformed into normal coordinates using the orthogonalizing matrix introduced in Eq.~(\ref{eq:eig_decomp}). It is assumed that the same matrix also orthogonalizes $\Lambda_{kl}$, such that
\begin{equation}
\sum_{l=1}^{N}\sum_{k=1}^{N}\Pi_{km}\Lambda_{kl}\Pi_{lq}=\widetilde{b}_{m}\delta_{mq}
\end{equation}
The quality of this assumption is found to be excellent, based on several test cases. The governing equation in normal coordinates is derived to be
\begin{equation}\label{eq:shear_norm_mode_diff_eq}
\begin{split}
\dfrac{d\bm{Q}'_{m}}{dt}&=\left(\dfrac{\widetilde{b}_m}{1+2\theta}\right)\boldsymbol{\kappa}\cdot\bm{Q}'_{m}-\left[\dfrac{H\widetilde{a}_m}{\zeta(1+2\theta)}\right]\bm{Q}'_{m}+\sqrt{\dfrac{2k_BT}{\zeta(1+2\theta)}}\sum_{k}\sum_{l}\Pi_{km}{B}_{kl}\,\bm{f}_{l}(t)
\end{split}
\end{equation}
Following a procedure identical to that described in Sec.~\ref{sec:derv_shear_flow}, the correlation between the Cartesian components of the normal modes can be derived to be
\begin{equation}
\begin{split}
\left<Q^{'(x)}_{p}(t)Q^{'(x)}_{q}(t)\right>&=\left(\dfrac{k_BT}{H}\right)\delta_{pq}+8\left(\dfrac{k_BT}{H}\right)\delta_{pq}\left(\dfrac{\widetilde{b}_q}{\widetilde{a}_q}\right)^2\left(\lambda_{\text{H}}\dot{\gamma}\right)^2\left[1-\left(e^{-2t/\widehat{\tau}_q}\left[1+\left(\dfrac{2t}{\widehat{\tau}_q}\right)\right]\right)\right]\\[5pt]
\left<Q^{'(y)}_{p}(t)Q^{'(y)}_{q}(t)\right>&=\left<Q^{'(z)}_{p}(t)Q^{'(z)}_{q}(t)\right>=\left(\dfrac{k_BT}{H}\right)\delta_{pq}\\[5pt]
\left<Q^{'(x)}_{p}(t)Q^{'(y)}_{q}(t)\right>&=\left(\dfrac{k_BT}{H}\right)\delta_{pq}\left(\lambda_{H}\dot{\gamma}\right)I_{q}(t)
\end{split}
\end{equation}
where
\begin{equation}
\begin{split}
I_{q}(t)&=2\left(\dfrac{\widetilde{b}_q}{\widetilde{a}_q}\right)\left[1-e^{-2t/\widehat{\tau}_q}\right].
\end{split}
\end{equation}
The equal-time correlation of the normal modes is subsequently obtained as
\begin{equation}\label{eq:correl_nmode_dot}
\begin{split}
&\left<\bm{Q}'_{p}(t)\cdot\bm{Q}'_{q}(t)\right>=\left(\dfrac{3k_BT}{H}\right)\delta_{pq}\Biggl\{1+\dfrac{8\left(\lambda_{\text{H}}\dot{\gamma}\right)^2}{3}\left(\dfrac{\widetilde{b}_p}{\widetilde{a}_p}\right)^2\left[1-\left(e^{-2t/\widehat{\tau}_p}\left[1+\left(\dfrac{2t}{\widehat{\tau}_p}\right)\right]\right)\right]\Biggr\},
\end{split}
\end{equation}
and the time evolution of the normalized mean-squared end-to-end distance may then be derived to be
\begin{equation}\label{eq:nmlzd_trans_re_t_preav}
\begin{split}
\dfrac{\left<\bm{R}_{\text{E}}^{*2}(t^*)\right>}{\left<\bm{R}^{*2}_{\text{E}}\right>_{\text{eq}}}=\left[\dfrac{1}{N_{\mathrm{b}}-1}\right]\,\sum_{i,j,p}^{N_{\text{b}}-1}\Pi_{ip}\Pi_{jp}&\Biggl\{1+\dfrac{8\left(\lambda_{\text{H}}\dot{\gamma}\right)^2}{3}\left(\dfrac{\widetilde{b}_p}{\widetilde{a}_p}\right)^2\Biggl[1-\Biggl(\exp\left[{-\left(\dfrac{\widetilde{a}_p}{1+2\theta}\right)\dfrac{t^{*}}{2}}\right]\\[5pt]
&\times\left[1+\left(\dfrac{\widetilde{a}_p}{1+2\theta}\right)\dfrac{t^{*}}{2}\right]\Biggr)\Biggr]\Biggr\}
\end{split}
\end{equation}
\end{widetext}

\begin{figure*}[t]
\begin{center}
\begin{tabular}{c c}
\includegraphics[width=9.25cm,height=!]{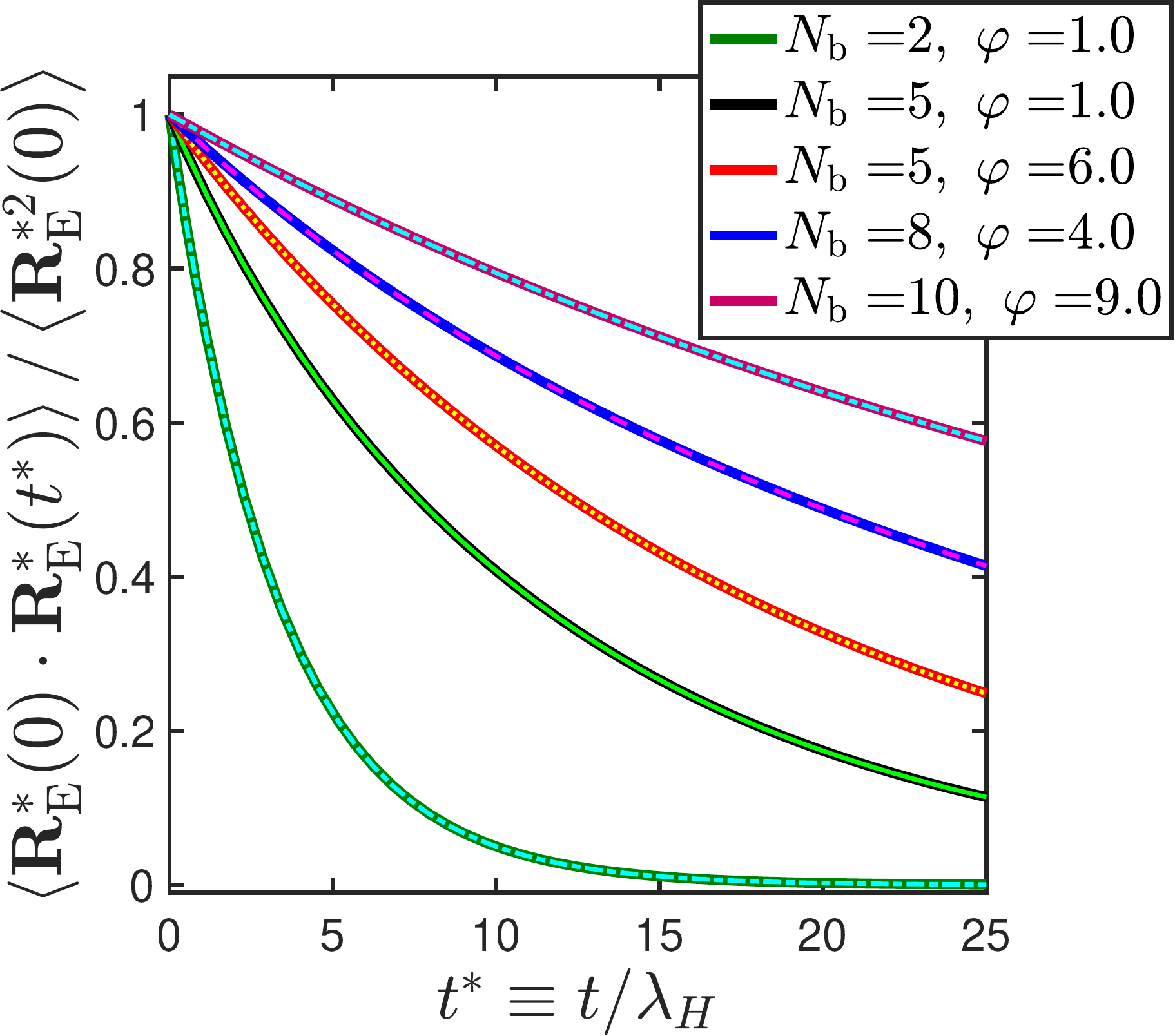}&
\includegraphics[width=8cm,height=!]{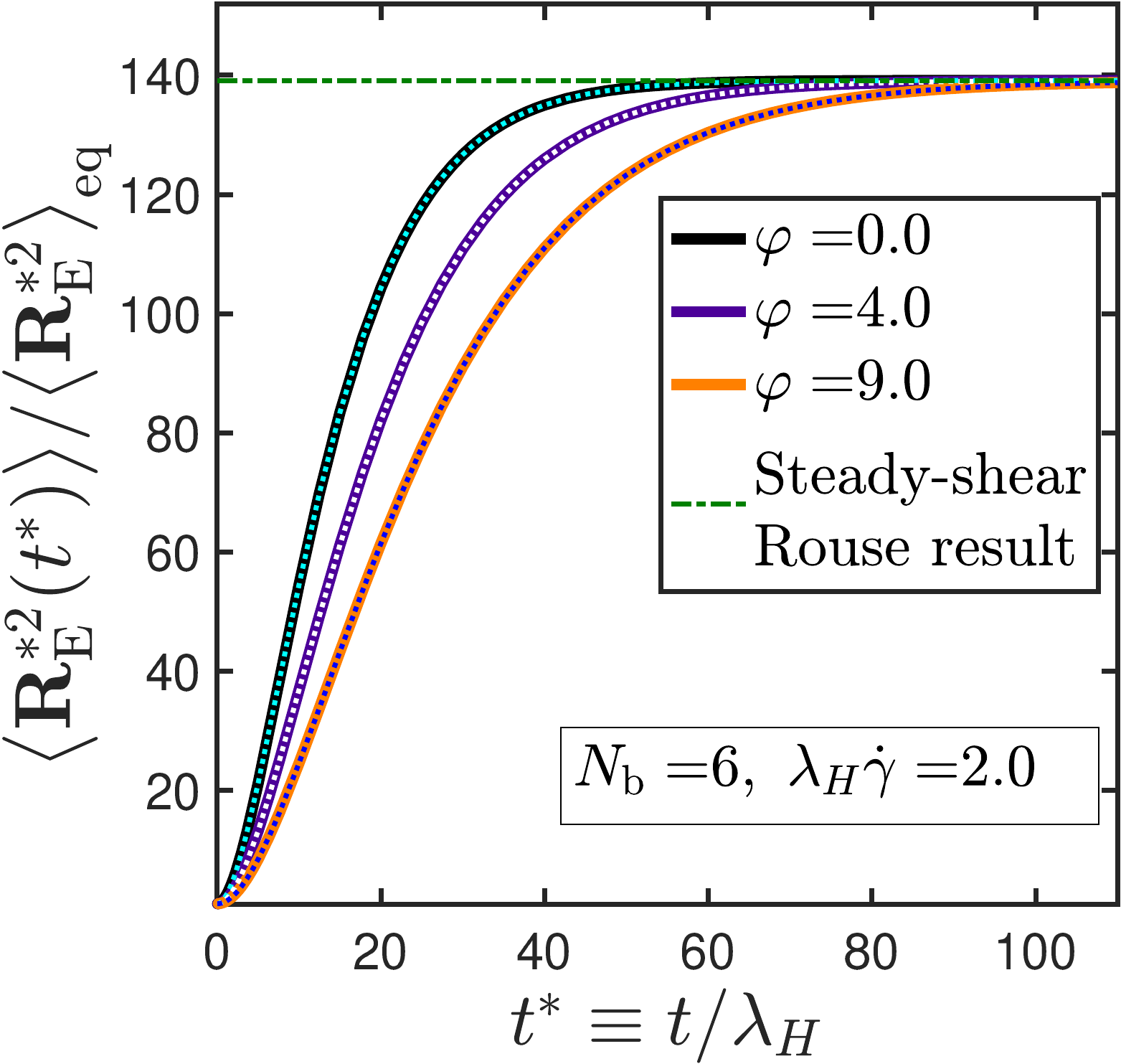}\\
(a) & (b)  \\
\end{tabular}
\end{center}
\caption{ A comparison of the results for the (a) normalized autocorrelation and (b) transient evolution of the mean-squared end-to-end vector predicted by the discrete RIF model and the preaveraged IV model derived using the principles of polymer kinetic theory. Each legend entry corresponds to two lines: the thicker lines represents the discrete RIF results [Eq.~(4) of the main text in (a) and Eq.~(\ref{eq:nmlzd_trans_re_t}) in (b)], while the thinner lines represents semi-analytical soutions for the preaveraged IV model [Eq.~\ref{eq:preav_disc_exp_nm_dimless} in (a) and Eq.~(\ref{eq:nmlzd_trans_re_t_preav}) in (b)].}
\label{fig:equivalence_drif_preav}
\end{figure*}

In Fig.~1 of the main text, the equivalence between the discrete RIF model and the preaveraged IV model is established by comparison of the analytical predictions of the discrete RIF model against BD simulation data obtained by numerically integrating the stochastic differential equation for the preaveraged IV model [Eq.~(\ref{eq:sde_individual})]. In Fig.~\ref{fig:equivalence_drif_preav}, predictions of the discrete RIF model are compared against the semi-analytical solutions for the preaveraged IV model, and an excellent agreement is observed.

We have therefore established that: (a) the discrete RIF model and the preaveraged IV model are equivalent, and (b) the preaveraged model may be solved for observables at equilibrium and in simple shear flow using either BD simulations to integrate the governing stochastic differential equation [Eq.~(\ref{eq:sde_individual})] or a semi-analytical approach involving the normal-mode decomposition of the governing Langevin equation [Eq.~(\ref{eq:langevin_general})].

\section{\label{sec:derv_stress_tensor} Derivation of stress tensor expression}

For models with fluctuating IV, it is known~\cite{Schieber1994} that while the Giesekus expression for the stress tensor is  thermodynamically consistent, the Kramers expression  is not. We have consequently used the Giesekus expression in our recent work~\cite{kailasham2021rouse} on Rouse chains with fluctuating internal friction. In the present instance, it has the form~\cite{Bird1987b}, 
\begin{widetext}
\begin{equation}\label{eq:giesekus_exp_start_point_preav}
\begin{split}
{\boldsymbol{\tau}_{\text{p}}}=\dfrac{n_{\text{p}}\zeta}{2}\left<\sum_{u=1}^{N}\sum_{v=1}^{N}{\mathscr{C}_{uv}}\boldsymbol{Q}_{u}\boldsymbol{Q}_{v}\right>_{(1)}=\dfrac{n_{\text{p}}\zeta}{2}\Biggl[&{\dfrac{d}{dt}\left<\sum_{u,v}{\mathscr{C}_{uv}}\boldsymbol{Q}_{u}\boldsymbol{Q}_{v}\right>}-\boldsymbol{\kappa}\cdot\left<\sum_{u,v}{\mathscr{C}_{uv}}\boldsymbol{Q}_{u}\boldsymbol{Q}_{v}\right>-\left<\sum_{u,v}{\mathscr{C}_{uv}}\boldsymbol{Q}_{u}\boldsymbol{Q}_{v}\right>\cdot\boldsymbol{\kappa}^{T}\Biggr]
\end{split}
\end{equation}
with $\mathscr{C}_{uv}$ denoting the Kramers matrix constructed as~\cite{Bird1987b}
 \begin{align}\label{eq:kramers_matrix_def}
\mathscr{C}_{uv}= \left\{
\begin{array}{ll}
       u(N_{\text{b}}-k)/N_{\text{b}}; &  u\leq v \\[15pt]
       v(N_{\text{b}}-u)/N_{\text{b}}; &  v\leq u 
\end{array} 
\right. 
\end{align}
We identify $\widetilde{\boldsymbol{B}}\equiv\sum_{u,v=1}^{N}{\mathscr{C}_{uv}}\boldsymbol{Q}_{u}\boldsymbol{Q}_{v}$ and note that a simplified, closed-form expression for the stress tensor in the present case may be found as illustrated below. Following the development outlined in Sec. 15.1 of Ref.~\citenum{Bird1987b}, the expression for the rate of change may be written as
\begin{equation}\label{eq:of_change_exp}
\dfrac{d}{dt}\left<\widetilde{\boldsymbol{B}}\right>=\left<\sum_{j=1}^{N}\left(\llbracket\dot{\boldsymbol{Q}}_{j}\rrbracket\cdot\dfrac{\partial \widetilde{\boldsymbol{B}}}{\partial \boldsymbol{Q}_{j}}\right)\right>
\end{equation}
and substitute into it the expression for $\llbracket\dot{\boldsymbol{Q}}_{j}\rrbracket$ given by Eq.~(B14) of Appendix B, to obtain
\begin{equation}\label{eq:db_dt_first_step}
\begin{split}
\dfrac{d}{dt}\left<\widetilde{\boldsymbol{B}}\right>&=\left(\dfrac{1}{1+2\theta}\right)\Biggl<\sum_{j,k}\Lambda_{jk}\left(\boldsymbol{\kappa}\cdot\boldsymbol{Q}_{k}\right)\cdot\dfrac{\partial \widetilde{\boldsymbol{B}}}{\partial \boldsymbol{Q}_{j}}\Biggr>-\left(\dfrac{k_BT}{\zeta}\right)\left(\dfrac{1}{1+2\theta}\right)\Biggl<\sum_{j,k}J_{jk}\dfrac{\partial \ln\psi}{\partial \boldsymbol{Q}_{k}}\cdot\dfrac{\partial \widetilde{\boldsymbol{B}}}{\partial \boldsymbol{Q}_{j}}\Biggr>\\[5pt]
&-\left(\dfrac{H}{\zeta}\right)\left(\dfrac{1}{1+2\theta}\right)\Biggl<\sum_{j,k}J_{jk}\boldsymbol{Q}_{k}\cdot\dfrac{\partial \widetilde{\boldsymbol{B}}}{\partial \boldsymbol{Q}_{j}}\Biggr>
\end{split}
\end{equation}
where we have used the choice of $\partial\phi/\partial \boldsymbol{Q}_{l}\equiv\,H\boldsymbol{Q}_{l}$ to restrict our attention to Rouse chains with Hookean springs. Also, the distribution function for the internal coordinates ($\psi$) has been used instead of $(\Psi)$, due to the homogeneous flow field under consideration. It is straightforward to derive the following expression for $\dfrac{\partial \widetilde{\boldsymbol{B}}}{\partial \boldsymbol{Q}_{j}}$,
\begin{equation}\label{eq:b_part}
\dfrac{\partial \widetilde{\boldsymbol{B}}}{\partial \boldsymbol{Q}_{j}}=\sum_{u}\sum_{\alpha,\beta}\mathscr{C}_{uj}Q^{\beta}_{u}\boldsymbol{e}_{\alpha}\boldsymbol{e}_{\beta}\boldsymbol{e}_{\alpha}+\sum_{v}\sum_{\alpha,\gamma}\mathscr{C}_{jv}Q^{\gamma}_{v}\boldsymbol{e}_{\alpha}\boldsymbol{e}_{\alpha}\boldsymbol{e}_{\gamma}
\end{equation}
The three terms on the RHS of Eq.~(\ref{eq:db_dt_first_step}) may then be processed as shown below.
\begin{equation}\label{eq:pf1}
\Biggl<\sum_{j,k}\Lambda_{jk}\left(\boldsymbol{\kappa}\cdot\boldsymbol{Q}_{k}\right)\cdot\dfrac{\partial \widetilde{\boldsymbol{B}}}{\partial \boldsymbol{Q}_{j}}\Biggr> = \Biggl<\sum_{k,u}\sum_{\beta,m,s}\left(\sum_{j}\Lambda_{kj}\mathscr{C}_{ju}\right)\kappa^{ms}Q^{s}_{k}Q^{\beta}_{u}\boldsymbol{e}_{\beta}\boldsymbol{e}_{m}\Biggr> + \Biggl<\sum_{k,v}\sum_{m,\gamma,s}\left(\sum_{j}\Lambda_{kj}\mathscr{C}_{jv}\right)\kappa^{ms}Q^{s}_{k}Q^{\gamma}_{v}\boldsymbol{e}_{m}\boldsymbol{e}_{\gamma}\Biggr>
\end{equation}
Defining the symmetric matrix $\boldsymbol{\mathcal{S}}=\boldsymbol{\Lambda}\cdot\boldsymbol{\mathcal{C}}$, i.e., $\mathcal{S}_{ku}=\sum_{j}\Lambda_{kj}\mathscr{C}_{ju}$, Eq.~(\ref{eq:pf1}) may be simplified to give
\begin{equation}\label{eq:flow_term_preav}
\begin{split}
\Biggl<\sum_{j,k}\Lambda_{jk}\left(\boldsymbol{\kappa}\cdot\boldsymbol{Q}_{k}\right)\cdot\dfrac{\partial \widetilde{\boldsymbol{B}}}{\partial \boldsymbol{Q}_{j}}\Biggr>&=\sum_{k,u}\mathcal{S}_{ku}\left[\left<\boldsymbol{Q}_u\boldsymbol{Q}_{k}\right>\cdot\boldsymbol{\kappa}^{T}+\boldsymbol{\kappa}\cdot\left<\boldsymbol{Q}_u\boldsymbol{Q}_{k}\right>^{T}\right]
\end{split}
\end{equation}
Next,
\begin{equation}\label{eq:pb1}
\begin{split}
\Biggl<\sum_{j,k}J_{jk}\dfrac{\partial \ln\psi}{\partial \boldsymbol{Q}_{k}}\cdot\dfrac{\partial \widetilde{\boldsymbol{B}}}{\partial \boldsymbol{Q}_{j}}\Biggr>&=\Biggl<\sum_{k,u}\sum_{\alpha,\beta}\left(\sum_{j}J_{kj}\mathscr{C}_{ju}\right)\dfrac{\partial \ln\psi}{\partial {Q}^{\alpha}_{k}}Q^{\beta}_{u}\boldsymbol{e}_{\beta}\boldsymbol{e}_{\alpha}\Biggr>+\Biggl<\sum_{k,v}\sum_{\alpha,\gamma}\left(\sum_{j}J_{kj}\mathscr{C}_{jv}\right)\dfrac{\partial \ln\psi}{\partial {Q}^{\alpha}_{k}}Q^{\gamma}_{v}\boldsymbol{e}_{\alpha}\boldsymbol{e}_{\gamma}\Biggr>
\end{split}
\end{equation}
\end{widetext}
Defining the symmetric matrix $\boldsymbol{\mathcal{L}}=\boldsymbol{J}\cdot\boldsymbol{\mathcal{C}}$, i.e., $\mathcal{L}_{ku}=\sum_{j}J_{kj}\mathscr{C}_{ju}$, the first term on the RHS of Eq.~(\ref{eq:pb1}) may be simplified to give
\begin{equation}
\begin{split}
\Biggl<\sum_{k,u}\sum_{\alpha,\beta}\mathcal{L}_{ku} & \dfrac{\partial \ln\psi}{\partial {Q}^{\alpha}_{k}}Q^{\beta}_{u}\boldsymbol{e}_{\beta}\boldsymbol{e}_{\alpha}\Biggr> = \\ 
& \sum_{k,u}\mathcal{L}_{ku}\sum_{\alpha,\beta}\int\left[\dfrac{\partial \psi}{\partial {Q}^{\alpha}_{k}}Q^{\beta}_{u}d\boldsymbol{Q}\right]\boldsymbol{e}_{\beta}\boldsymbol{e}_{\alpha}\\[5pt]
&=-\sum_{k,u}\mathcal{L}_{ku}\delta_{ku}\sum_{\alpha,\beta}\delta^{\alpha\beta}\boldsymbol{e}_{\beta}\boldsymbol{e}_{\alpha}\\[5pt]
&=-\text{tr}\left(\boldsymbol{\mathcal{L}}\right)\boldsymbol{\delta}
\end{split}
\end{equation}
Therefore, 
\begin{equation}\label{eq:brown_preav}
\Biggl<\sum_{j,k}J_{jk}\dfrac{\partial \ln\psi}{\partial \boldsymbol{Q}_{k}}\cdot\dfrac{\partial \widetilde{\boldsymbol{B}}}{\partial \boldsymbol{Q}_{j}}\Biggr>  =  
-2\text{tr}\left(\boldsymbol{\mathcal{L}}\right)\boldsymbol{\delta}
\end{equation}
Lastly, the third term on the RHS of Eq.~(\ref{eq:db_dt_first_step}) may be simplified as
\begin{equation}\label{eq:pforce1}
\begin{split}
\Biggl<\sum_{j,k} & J_{jk}\boldsymbol{Q}_{k}\cdot\dfrac{\partial \widetilde{\boldsymbol{B}}}{\partial \boldsymbol{Q}_{j}}\Biggr>
=\Biggl<\sum_{k,u}\sum_{\alpha,\beta}\left(\sum_{j}J_{kj}\mathscr{C}_{ju}\right)Q^{\alpha}_{k}Q^{\beta}_{u}\boldsymbol{e}_{\beta}\boldsymbol{e}_{\alpha}\Biggr> \\ 
& +\Biggl<\sum_{k,v}\sum_{\alpha,\gamma}\left(\sum_{j}J_{kj}\mathscr{C}_{jv}\right)Q^{\alpha}_{k}Q^{\gamma}_{v}\boldsymbol{e}_{\alpha}\boldsymbol{e}_{\gamma}\Biggr>\\[5pt]
&=\sum_{k,u}\mathcal{L}_{ku}\Biggl[\left<\boldsymbol{Q}_{k}\boldsymbol{Q}_{u}\right>+\left<\boldsymbol{Q}_{k}\boldsymbol{Q}_{u}\right>^{T}\Biggr]
\end{split}
\end{equation}

From Equations~(\ref{eq:giesekus_exp_start_point_preav}), ~(\ref{eq:flow_term_preav}), ~(\ref{eq:brown_preav}), and ~(\ref{eq:pforce1}), we have
\begin{widetext}
\begin{equation}\label{eq:lower_conv_preav}
\begin{split}
&\left<\sum_{u=1}^{N}\sum_{v=1}^{N}{\mathscr{C}_{uv}}\boldsymbol{Q}_{u}\boldsymbol{Q}_{v}\right>_{(1)}=\left(\dfrac{1}{1+2\theta}\right)\sum_{k,u}\mathcal{S}_{ku}\left[\left<\boldsymbol{Q}_u\boldsymbol{Q}_{k}\right>\cdot\boldsymbol{\kappa}^{T}+\boldsymbol{\kappa}\cdot\left<\boldsymbol{Q}_u\boldsymbol{Q}_{k}\right>^{T}\right]+\left(\dfrac{2k_BT}{\zeta}\right)\left(\dfrac{1}{1+2\theta}\right)\text{tr}\left(\boldsymbol{\mathcal{L}}\right)\\[5pt]
&+\sum_{k,u}\mathcal{L}_{ku}\Biggl[\left<\boldsymbol{Q}_{k}\boldsymbol{Q}_{u}\right>+\left<\boldsymbol{Q}_{k}\boldsymbol{Q}_{u}\right>^{T}\Biggr]\boldsymbol{\delta}-\boldsymbol{\kappa}\cdot\left<\sum_{u,v}{\mathscr{C}_{uv}}\boldsymbol{Q}_{u}\boldsymbol{Q}_{v}\right>-\left<\sum_{u,v}{\mathscr{C}_{uv}}\boldsymbol{Q}_{u}\boldsymbol{Q}_{v}\right>\cdot\boldsymbol{\kappa}^{T}
\end{split}
\end{equation}
Substituting Eq.~(\ref{eq:lower_conv_preav}) into Eq.~(\ref{eq:giesekus_exp_start_point_preav}) and simplifying, the following expression for the stress tensor is obtained
\begin{equation}\label{eq:preav_dim}
\begin{split}
\boldsymbol{\tau}_{\text{p}}&=n_{\text{p}}k_BT\left[\left(\dfrac{1}{1+2\theta}\right)\text{tr}\left(\bm{\mathcal{L}}\right)\right]\boldsymbol{\delta}-\dfrac{n_{\text{p}}H}{2}\left[\left(\dfrac{1}{1+2\theta}\right)\sum_{m,n}\mathcal{L}_{mn}\left[\left<\bm{Q}_{m}\bm{Q}_n\right>+\left<\bm{Q}_n\bm{Q}_{m}\right>\right]\right]\\[5pt]
&-\dfrac{n_{\text{p}}\zeta}{2}\Biggl\{\sum_{m,n}\left(\mathscr{C}_{mn}-\dfrac{1}{1+2\theta}\mathcal{S}_{mn}\right)\left[\left<\boldsymbol{\kappa}\cdot\left(\bm{Q}_m\bm{Q}_n\right)\right>+\left<\left(\bm{Q}_m\bm{Q}_n\right)\cdot\boldsymbol{\kappa}^{T}\right>\right]\Biggr\}
\end{split}
\end{equation}
\end{widetext}
The $xy-$ component of the stress tensor is written as
\begin{equation}
\begin{split}
\tau_{\text{p},xy}&=-{n_{\text{p}}H}\left[\left(\dfrac{1}{1+2\theta}\right)\sum_{m,n}\mathcal{L}_{mn}\dashuline{\left<{Q}^{(x)}_{m}{Q}^{(y)}_n\right>}\right] \\
&- \dfrac{n_{\text{p}}\zeta\dot{\gamma}}{2}\Biggl\{\sum_{m,n}\left(\mathscr{C}_{mn}-\dfrac{1}{1+2\theta}\mathcal{S}_{mn}\right)\left[\dotuline{\left<{Q}^{(y)}_m{Q}^{(y)}_n\right>}\right]\Biggr\}
\end{split}
\end{equation}

\begin{figure}[t]
\centering
\includegraphics[width=8cm,height=!]{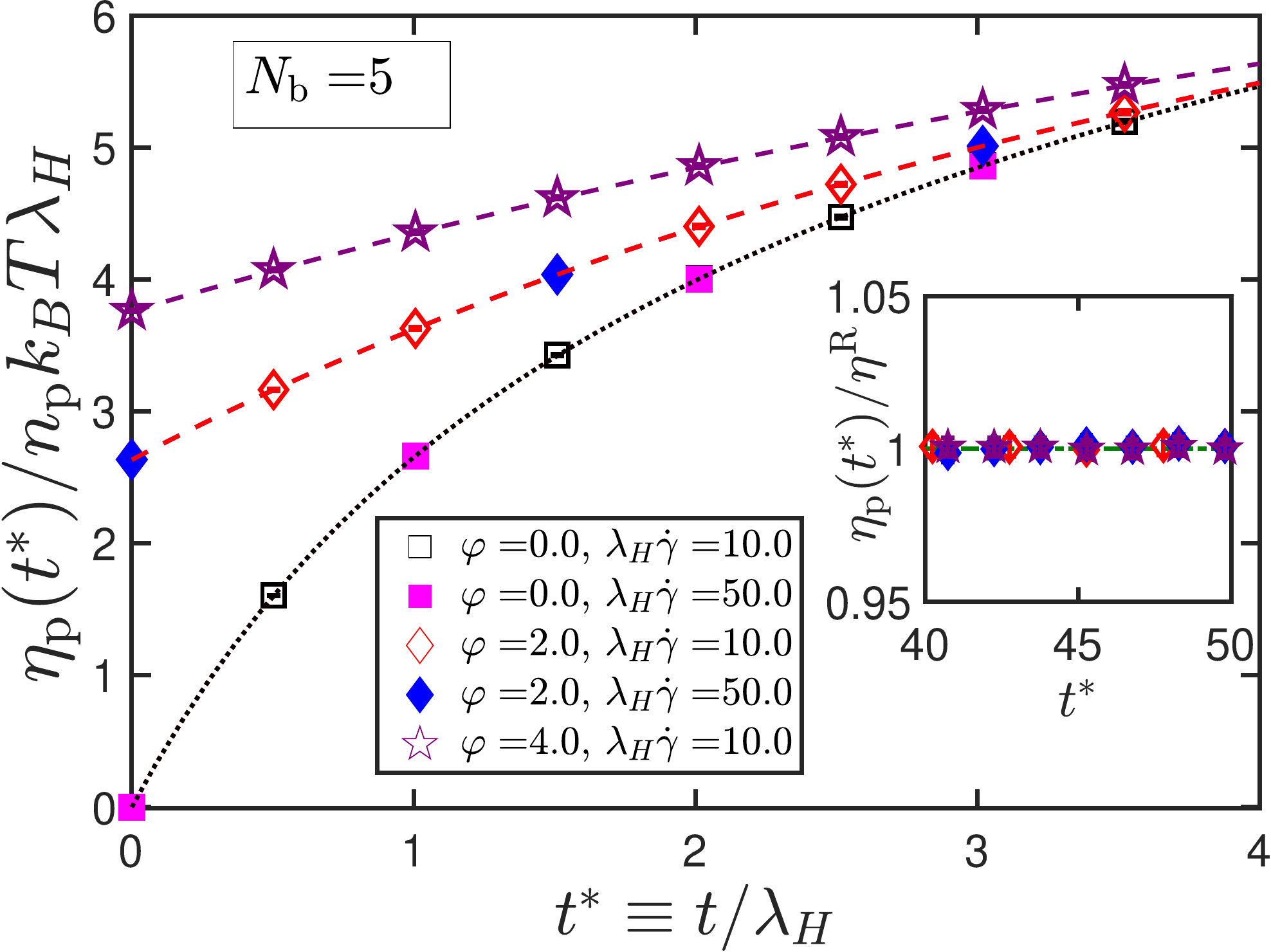}  
\caption{\small Validating the expression for the transient evolution of shear viscosity for a five-bead Rouse chain with preaveraged internal friction for various values of the dimensionless shear rate ($\lambda_{H}\dot{\gamma}$) and internal friction parameter ($\varphi$). Lines represent Eq.~(\ref{eq:preav_iv_trans_eta}) and symbols are BD simulations results. Inset shows the steady-state values of viscosity for the various cases.}
\label{fig:shear_compare2}
\end{figure}

Following a procedure identical to that described in Sec.~\ref{sec:derv_shear_flow}, the underlined terms are evaluated to be
\begin{equation}
\begin{split}
\left<{Q}^{(x)}_{m}{Q}^{(y)}_n\right>&=\left(\dfrac{k_BT}{H}\right)\left(\lambda_{H}\dot{\gamma}\right)\sum_{q}\Pi_{mq}\Pi_{nq}I_{q}(t)\\[5pt]
\left<{Q}^{(y)}_m{Q}^{(y)}_n\right>&=\left(\dfrac{k_BT}{H}\right)\delta_{mn}
\end{split}
\end{equation}
from which the expression for the dimensionless shear viscosity may be obtained as
\begin{equation}\label{eq:preav_iv_trans_eta2}
\begin{split}
\dfrac{\eta_{\text{p}}(t^*)}{n_{\text{p}}k_BT\lambda_{H}}&\equiv-\dfrac{\tau_{\text{p},xy}}{n_{\text{p}}k_BT\lambda_{H}\dot{\gamma}}\\[5pt]
&=\left(\dfrac{1}{1+2\theta}\right)\sum_{m,n,q}\Pi_{mq}\mathcal{L}_{mn}\Pi_{nq}I_{q}(t^*) \\
& + 2\,\text{tr}\left[\bm{\mathcal{C}}-\dfrac{1}{1+2\theta}\bm{\mathcal{S}}\right]
\end{split}
\end{equation}
with
\begin{equation}
\begin{split}
I_{q}(t^*)&=2\left(\dfrac{\widetilde{b}_q}{\widetilde{a}_q}\right)\left\{1-\exp\left[-\left(\dfrac{\widetilde{a}_q}{1+2\theta}\right)\dfrac{t^{*}}{2}\right]\right\}
\end{split}
\end{equation}
Equation~(\ref{eq:preav_iv_trans_eta2}) is reproduced as Eq.~(\ref{eq:preav_iv_trans_eta}) in the main text, with $\varphi$ in place of $\theta$.

In Fig.~\ref{fig:shear_compare2}, the solution given by Eq.~(\ref{eq:preav_iv_trans_eta}) is compared against BD simulation data for the preaveraged IV model, obtained by numerically integrating Eq.~(\ref{eq:sde_individual}) and using the stress tensor expression given by Eq.~(\ref{eq:preav_dim}), for a variety of internal friction parameters and shear rates. An excellent agreement is observed between the results obtained using the two approaches.

%




\end{document}